\documentclass[10pt,journal,compsoc]{IEEEtran}

\usepackage{array,multirow}
\usepackage{booktabs} 
\usepackage{graphicx}
\usepackage{caption}
\usepackage{subcaption}
\usepackage{amssymb}
\usepackage{threeparttable}
\usepackage{booktabs}
\usepackage[nocompress]{cite}
\usepackage{bm}
\usepackage{amsmath,amssymb,amsfonts}
\usepackage{algorithmic}
\usepackage{graphicx}
\usepackage{mathtools}
\usepackage{textcomp}
\usepackage{xcolor}
\usepackage{physics}
\usepackage[hyphens]{url}

\usepackage[capitalize]{cleveref}
\crefname{section}{Sec.}{Secs.}

\def\BibTeX{{\rm B\kern-.05em{\sc i\kern-.025em b}\kern-.08em
    T\kern-.1667em\lower.7ex\hbox{E}\kern-.125emX}}

\usepackage[acronyms,nonumberlist,nopostdot,nomain,nogroupskip]{glossaries}
\newacronym{sla}{SLA}{Service Level Agreement}
\newacronym{urllc}{URLLC}{Ultra Reliable Low Latency Communication}
\newacronym{mmtc}{mMTC}{massive Machine Type Communication}
\newacronym{sdn}{SDN}{Software Defined Networking}
\newacronym{ns}{NS}{Network Slicing}
\newacronym{bbu}{BBU}{BaseBand Unit}
\newacronym{psc}{PSC}{Public Safety Communication}
\newacronym{lmr}{LMR}{Land Mobile Radio}
\newacronym{bb}{BB}{Broadband}
\newacronym{ps}{PS}{Public Safety}
\newacronym{uav}{UAV}{Unmaned Aerial Vehicle}
\newacronym{arma}{ARMA}{Auto-Regressive Moving Average}
\newacronym{lstm}{LSTM}{Long Short-Term Memory}

\newacronym{marl}{MARL}{Multi-Agent Reinforcement Learning}
\newacronym{drl}{DRL}{Deep Reinforcement Learning}
\newacronym{pomdp}{POMDP}{Partially Observable Markov Decision Process}
\newacronym{mdp}{MDP}{Markov Decision Process}
\newacronym{cnn}{CNN}{Convolutional Neural Network}
\newacronym{nn}{NN}{Neural Network}
\newacronym{radam}{RAdam}{Rectified Adam}
\newacronym{relu}{ReLU}{Rectified Linear Unit}
\newacronym{adam}{Adam}{Adaptive moment estimator}
\newacronym{ac}{AC}{Actor Critic}
\newacronym{a2c}{A2C}{Advance Actor Critic}
\newacronym{gnn}{GNN}{Graph Neural Network}
\newacronym{fl}{FL}{Federated Learning}
\newacronym{sarsa}{SARSA}{State-Action-Reward-State-Action}
\newacronym{mc}{MC}{Markov Chain}

\newacronym{fifo}{FIFO}{First In First Out}
\newacronym{cvo}{CVO}{Critical Voice}
\newacronym{cvi}{CVI}{Critical Video}
\newacronym{ncvo}{NCVO}{Non-Critical Voice}
\newacronym{ncvi}{NCVI}{Non-Critical Video}
\newacronym{ma}{MA}{Moving Average}

\newacronym{3dof}{3DoF}{3 Degrees of Freedom}
\newacronym{3gpp}{3GPP}{3rd Generation Partnership Project}
\newacronym{5g}{5G}{5\textsuperscript{th} Generation}
\newacronym{5gc}{5GC}{5G Core}
\newacronym{6dof}{6DoF}{6 Degrees of Freedom}
\newacronym{abft}{A-BFT}{Association-BeamForming Training}
\newacronym{adc}{ADC}{Analog to Digital Converter}
\newacronym{addts}{ADDTS}{Add Traffic Stream}
\newacronym{afbw}{AFBW}{Average Fading Bandwidth}
\newacronym{aid}{AID}{Association ID}
\newacronym{aifs}{AIFS}{Arbitration Inter-Frame Space}
\newacronym{aimd}{AIMD}{Additive Increase Multiplicative Decrease}
\newacronym{am}{AM}{Acknowledged Mode}
\newacronym{amc}{AMC}{Adaptive Modulation and Coding}
\newacronym{ampdu}{A-MPDU}{MAC Protocol Data Unit Aggregation}
\newacronym{aoa}{AoA}{Angle of Arrival}
\newacronym{aod}{AoD}{Angle of Departure}
\newacronym{ap}{AP}{Access Point}
\newacronym{api}{API}{Application Programming Interface}
\newacronym{app}{APP}{Application}
\newacronym{aqm}{AQM}{Active Queue Management}
\newacronym{ar}{AR}{Augmented Reality}
\newacronym{arf}{ARF}{Auto Rate Fallback}
\newacronym{arp}{ARP}{Address Resolution Protocol}
\newacronym{ati}{ATI}{Announcement Transmission Interval}
\newacronym{awgn}{AGWN}{Additive White Gaussian Noise}
\newacronym{awv}{AWV}{Antenna Weight Vector}
\newacronym{balia}{BALIA}{Balanced Link Adaptation}
\newacronym{bdp}{BDP}{Bandwidth-Delay Product}
\newacronym{ber}{BER}{Bit Error Rate}
\newacronym{bf}{BF}{Beamforming}
\newacronym{bframe}{B-frame}{Bipredictive-coded frame}
\newacronym{bhi}{BHI}{Beacon Header Interval}
\newacronym{bi}{BI}{Beacon Interval}
\newacronym{brp}{BRP}{Beam Refinement Protocol}
\newacronym{bs}{BS}{Base Station}
\newacronym{bss}{BSS}{Basic Service Set}
\newacronym{bti}{BTI}{Beacon Transmission Interval}
\newacronym{cad}{CAD}{Computer-aided Design}
\newacronym{cbap}{CBAP}{Contention-Based Access Period}
\newacronym{cbr}{CBR}{Constant Bit Rate}
\newacronym{cc}{CC}{Congestion Control}
\newacronym{cdf}{CDF}{Cumulative Distribution Function}
\newacronym{cir}{CIR}{Channel Impulse Response}
\newacronym{cn}{CN}{Core Network}
\newacronym{cp}{CP}{Control Plane}
\newacronym{cqi}{CQI}{Channel Quality Indicator}
\newacronym{crs}{CRS}{Cell Reference Signal}
\newacronym{csirs}{CSI-RS}{Channel State Information - Reference Signal}
\newacronym{csmaca}{CSMA/CA}{Carrier Sense Multiple Access with Collision Avoidance}
\newacronym{cts}{CTS}{Clear to Send}
\newacronym{dc}{DC}{Dual Connectivity}
\newacronym{dce}{DCE}{Direct Code Execution}
\newacronym{dcf}{DCF}{Distributed Coordination Function}
\newacronym{dci}{DCI}{Downlink Control Information}
\newacronym{delts}{DELTS}{Delete Traffic Stream}
\newacronym{dl}{DL}{Downlink}
\newacronym{dmg}{DMG}{Directional Multi-Gigabit}
\newacronym{dmr}{DMR}{Deadline Miss Ratio}
\newacronym{dmrs}{DMRS}{DeModulation Reference Signal}
\newacronym{dti}{DTI}{Data Transmission Interval}
\newacronym{e2e}{E2E}{End-to-End}
\newacronym{ecn}{ECN}{Explicit Congestion Notification}
\newacronym{edca}{EDCA}{Enhanced Distributed Channel Access}
\newacronym{edf}{EDF}{Earliest Deadline First}
\newacronym{embb}{eMBB}{Enhanced Mobile BroadBand}
\newacronym{enb}{eNB}{evolved Node Base}
\newacronym{endc}{EN-DC}{E-UTRAN-\gls{nr} \gls{dc}}
\newacronym{epc}{EPC}{Evolved Packet Core}
\newacronym{es}{ES}{Edge Server}
\newacronym{ese}{ESE}{Extended Schedule Element}
\newacronym{fdd}{FDD}{Frequency Division Duplexing}
\newacronym{fdma}{FDMA}{Frequency Division Multiple Access}
\newacronym{fec}{FEC}{Forward Error Correction}
\newacronym{fov}{FoV}{Field-of-View}
\newacronym{fps}{FPS}{Frames per Second}
\newacronym{fr2}{FR2}{Frequency Range 2}
\newacronym{ftp}{FTP}{File Transfer Protocol}
\newacronym{gmm}{GMM}{Gaussian Mixture Model}
\newacronym{gnb}{gNB}{Next Generation Node Base}
\newacronym[firstplural=Group of Pictures (GoPs)]{gop}{GoP}{Group of Pictures}
\newacronym{harq}{HARQ}{Hybrid Automatic Repeat reQuest}
\newacronym{hetnet}{HetNet}{Heterogeneous Network}
\newacronym{hh}{HH}{Hard Handover}
\newacronym{hmd}{HMD}{Head Mounted Device}
\newacronym{hol}{HOL}{Head-of-Line}
\newacronym{hqf}{HQF}{Highest-quality-first}
\newacronym{ia}{IA}{Initial Access}
\newacronym{iab}{IAB}{Integrated Access and Backhaul}
\newacronym{ibss}{IBSS}{Independent Basic Service Set}
\newacronym{id}{ID}{Identifier}
\newacronym{ifi}{IFI}{Inter-Frame Inter-arrival}
\newacronym{iframe}{I-frame}{Intra-coded frame}
\newacronym{imt}{IMT}{International Mobile Telecommunication}
\newacronym{imt2020}{IMT-2020}{International Mobile Te\-le\-com\-mu\-ni\-ca\-tion-2020}
\newacronym{inr}{INR}{Interference to Noise Ratio}
\newacronym{iot}{IoT}{Internet of Things}
\newacronym{ipa}{IPA}{Inter-Packet Arrival}
\newacronym{ism}{ISM}{Industrial, Scientific, and Medical}
\newacronym{itu}{ITU}{International Telecommunication Union}
\newacronym{kpi}{KPI}{Key Performance Indicator}
\newacronym{ks}{KS}{Kolmogorov-Smirnov}
\newacronym{lcf}{LCF}{Level Crossing Frequency}
\newacronym{lcm}{lcm}{least common multiple}
\newacronym{lcr}{LCR}{Level Crossing Rate}
\newacronym{los}{LoS}{Line-of-Sight}
\newacronym{lp}{LP}{Low Power}
\newacronym{lte}{LTE}{Long Term Evolution}
\newacronym{m2m}{M2M}{Machine to Machine}
\newacronym{ols}{OLS}{Ordinary Least Squares}
\newacronym{mac}{MAC}{Medium Access Control}
\newacronym{mcs}{MCS}{Modulation and Coding Scheme}
\newacronym{mec}{MEC}{Mobile Edge Cloud}
\newacronym{mi}{MI}{Mutual Information}
\newacronym{mib}{MIB}{Master Information Block}
\newacronym{mimo}{MIMO}{Multiple Input, Multiple Output}
\newacronym{ml}{ML}{Machine Learning}
\newacronym{mlr}{MLR}{Maximum-local-rate}
\newacronym[plural=\gls{mme}s,firstplural=Mobility Management Entities (MMEs)]{mme}{MME}{Mobility Management Entity}
\newacronym{mmw}{mmW}{Millimeter Wave}
\newacronym{moi}{MoI}{Method of Images}
\newacronym{mtp}{MTP}{Motion-To-Photon}
\newacronym{mpc}{MPC}{Multi Path Component}
\newacronym{mpdu}{MPDU}{MAC Protocol Data Unit}
\newacronym{mptcp}{MPTCP}{Multipath TCP}
\newacronym{mr}{MR}{Mixed Reality}
\newacronym{mrdc}{MR-DC}{Multi \gls{rat} \gls{dc}}
\newacronym{msdu}{MSDU}{MAC Service Data Unit}
\newacronym{mss}{MSS}{Maximum Segment Size}
\newacronym{mtd}{MTD}{Machine-Type Device}
\newacronym{mtu}{MTU}{Maximum Transmission Unit}
\newacronym{mumimo}{MU-MIMO}{Multi-User Multiple Input, Multiple Output}
\newacronym{nav}{NAV}{Network Allocation Vector}
\newacronym{ncbr}{NCBR}{Non-Constant Bit Rate}
\newacronym{nfv}{NFV}{Network Function Virtualization}
\newacronym{nlos}{NLoS}{Non-Line-of-Sight}
\newacronym{nr}{NR}{New Radio}
\newacronym{mcmc}{MCMC}{Markov Chain Monte Carlo}
\newacronym{mse}{MSE}{Mean Square Error}
\newacronym{nrmse}{NRMSE}{Normalized Root Mean Square Error}
\newacronym{ns2}{ns-2}{Network Simulator 2}
\newacronym{ns3}{ns-3}{Network Simulator 3}
\newacronym{nsa}{NSA}{Non Stand Alone}
\newacronym{o2i}{O2I}{Outdoor-to-Indoor}
\newacronym{ofdma}{OFDMA}{Orthogonal Frequency Division Multiple Access}
\newacronym{pa}{PA}{Position-aware}
\newacronym{pan}{PAN}{Personal Area Network}
\newacronym{pbch}{PBCH}{Physical Broadcast Channel}
\newacronym{pbss}{PBSS}{Personal Basic Service Set}
\newacronym{pcf}{PCF}{Point Coordinator Function}
\newacronym{pcp}{PCP}{\gls{pbss} Central Point}
\newacronym{pcpap}{PCP/AP}{\acrlong{pcp}/\acrlong{ap}}
\newacronym{pdcch}{PDCCH}{Physical Downlonk Control Channel}
\newacronym{pdcp}{PDCP}{Packet Data Convergence Protocol}
\newacronym{pdf}{PDF}{Probability Density Function}
\newacronym{pdsch}{PDSCH}{Physical Downlink Shared Channel}
\newacronym{pdu}{PDU}{Packet Data Unit}
\newacronym{per}{PER}{Packet Error Rate}
\newacronym{pf}{PF}{Proportional Fair}
\newacronym{pframe}{P-frame}{Predictive-coded frame}
\newacronym{pgw}{PGW}{Packet Gateway}
\newacronym{phy}{PHY}{Physical Layer}
\newacronym{ppdu}{PPDU}{PHY Protocol Data Unit}
\newacronym{ppp}{PPP}{Poisson Point Process}
\newacronym{prb}{PRB}{Physical Resource Block}
\newacronym{pss}{PSS}{Primary Synchronization Signal}
\newacronym{pucch}{PUCCH}{Physical Uplink Control Channel}
\newacronym{pusch}{PUSCH}{Physical Uplink Shared Channel}
\newacronym{qd}{QD}{Quasi Deterministic}
\newacronym{qoe}{QoE}{Quality of Experience}
\newacronym{qos}{QoS}{Quality of Service}
\newacronym{rach}{RACH}{Random Access Channel}
\newacronym{ran}{RAN}{Radio Access Network}
\newacronym[firstplural=Radio Access Technologies (RATs)]{rat}{RAT}{Radio Access Technology}
\newacronym{red}{RED}{Random Early Detection}
\newacronym{rf}{RF}{Radio Frequency}
\newacronym{rl}{RL}{Reinforcement Learning}
\newacronym{rlc}{RLC}{Radio Link Control}
\newacronym{rlf}{RLF}{Radio Link Failure}
\newacronym{rr}{RR}{Round Robin}
\newacronym{rrc}{RRC}{Radio Resource Control}
\newacronym{rrm}{RRM}{Radio Resource Management}
\newacronym{rs}{RS}{Remote Server}
\newacronym{rsrp}{RSRP}{Reference Signal Received Power}
\newacronym{rsrq}{RSRQ}{Reference Signal Received Quality}
\newacronym{rss}{RSS}{Received Signal Strength}
\newacronym{rssi}{RSSI}{Received Signal Strength Indicator}
\newacronym{rt}{RT}{Ray Tracer}
\newacronym{rts}{RTS}{Request to Send}
\newacronym{rtt}{RTT}{Round Trip Time}
\newacronym{rw}{RW}{Receive Window}
\newacronym{rx}{RX}{Receiver}
\newacronym{sa}{SA}{standalone}
\newacronym{sack}{SACK}{Selective Acknowledgment}
\newacronym{sap}{SAP}{Service Access Point}
\newacronym{sc}{SC}{Single Carrier}
\newacronym{sch}{SCH}{Secondary Cell Handover}
\newacronym{scm}{SCM}{Spatial Channel Model}
\newacronym{scoot}{SCOOT}{Split Cycle Offset Optimization Technique}
\newacronym{sdma}{SDMA}{Spatial Division Multiple Access}
\newacronym{sdr}{SDR}{Software Defined Radio}
\newacronym{semm}{SEMM}{SPCA-EDCA Mixed Mode}
\newacronym{si}{SI}{Study Item}
\newacronym{sib}{SIB}{Secondary Information Block}
\newacronym{sinr}{SINR}{Signal-to-Interference-plus-Noise Ratio}
\newacronym{sir}{SIR}{Signal-to-Interference Ratio}
\newacronym{sls}{SLS}{Sector-Level Sweep}
\newacronym{sm}{SM}{Saturation Mode}
\newacronym{snr}{SNR}{Signal-to-Noise Ratio}
\newacronym{son}{SON}{Self-Organizing Network}
\newacronym{sp}{SP}{Service Period}
\newacronym{spr}{SPR}{Service Period Request}
\newacronym{srs}{SRS}{Sounding Reference Signal}
\newacronym{ss}{SS}{Synchronization Signal}
\newacronym{sss}{SSS}{Secondary Synchronization Signal}
\newacronym{ssw}{SSW}{Sector Sweep}
\newacronym{sta}{STA}{Station}
\newacronym{stb}{STB}{Set Top Box}
\newacronym{tb}{TB}{Transport Block}
\newacronym{tbtt}{TBTT}{Target Beacon Transmission Time}
\newacronym[firstplural=Traffic Categories (TCs)]{tc}{TC}{Traffic Category}
\newacronym{tcp}{TCP}{Transmission Control Protocol}
\newacronym{tdd}{TDD}{Time Division Duplexing}
\newacronym{tdma}{TDMA}{Time Division Multiple Access}
\newacronym{tfl}{TfL}{Transport for London}
\newacronym{tgad}{TGad}{Task Group ad}
\newacronym{tgay}{TGay}{Task Group ay}
\newacronym{tm}{TM}{Transparent Mode}
\newacronym{trp}{TRP}{Transmitter Receiver Pair}
\newacronym{ts}{TS}{Traffic Stream}
\newacronym{tsconst}{TSCONST}{Traffic Scheduling Constraint}
\newacronym{tsf}{TSF}{Timing Synchronization Function}
\newacronym{tspec}{TSPEC}{Traffic Specification}
\newacronym{tti}{TTI}{Transmission Time Interval}
\newacronym{ttt}{TTT}{Time-to-Trigger}
\newacronym{tx}{TX}{Transmitter}
\newacronym[firstplural=Transmission Opportunities (TXOPs)]{txop}{TXOP}{Transmission Opportunity}
\newacronym{udp}{UDP}{User Datagram Protocol}
\newacronym{ue}{UE}{User Equipment}
\newacronym{ul}{UL}{Uplink}
\newacronym{um}{UM}{Unacknowledged Mode}
\newacronym{uma}{UMa}{Urban Macro}
\newacronym{uml}{UML}{Unified Modeling Language}
\newacronym{up}{UP}{User Priority}
\newacronym{utc}{UTC}{Urban Traffic Control}
\newacronym{vbr}{VBR}{Variable Bit Rate}
\newacronym{vm}{VM}{Virtual Machine}
\newacronym{vr}{VR}{Virtual Reality}
\newacronym{wbf}{WBF}{Wired Bias Function}
\newacronym{wf}{WF}{Wired-first}
\newacronym{wifi}{Wi-Fi}{Wireless Fidelity}
\newacronym{wigig}{WiGig}{Wireless Gigabit}
\newacronym{wlan}{WLAN}{Wireless Local Area Network}
\newacronym{xr}{XR}{eXtended Reality}

\newacronym{nvenc}{NVENC}{Nvidia Encoder}

\newacronym{gm}{GM}{\emph{general} model}
\newacronym{cm}{CM}{\emph{content-dependent} model}
\newacronym{crm}{CRM}{\emph{content- and rate-dependent} model}
\newacronym{fs}{FS}{\emph{Frame-by-frame} scheduling}
\newacronym{cs}{CS}{\emph{Constant} scheduling}

\newacronym{if}{IF}{Individual FDMA}
\newacronym{af}{AF}{Aggregated FDMA}
\newacronym{io}{IO}{Individual OFDMA}
\newacronym{ao}{AO}{Aggregated OFDMA}

\newacronym{ieee}{IEEE}{Institute of Electrical and Electronics Engineers}

\usepackage{tikz}
\usepackage{pgfplots}
\usepackage{tikzscale}
\usepackage{tikz-qtree}
\pgfplotsset{compat=newest}
\pgfplotsset{plot coordinates/math parser=false}
\pgfplotsset{every axis/.append style={
                    label style={font=\scriptsize},
                    tick label style={font=\scriptsize},
                    legend style={font=\scriptsize}
                    }}
\usetikzlibrary{plotmarks,shapes,patterns,decorations.pathreplacing,backgrounds,calc,arrows,arrows.meta,spy,matrix,shadows,trees,positioning,fit}
\usepgfplotslibrary{patchplots,groupplots}

\tikzstyle{startstop} = [rectangle, rounded corners, minimum width=2cm, minimum height=0.5cm,text centered, draw=black]
\tikzstyle{io} = [trapezium, trapezium left angle=70, trapezium right angle=110, minimum width=3cm, minimum height=1cm, text centered, draw=black]
\tikzstyle{process} = [rectangle, minimum width=2cm, minimum height=0.5cm, text centered, draw=black, alignb=center]
\tikzstyle{decision} = [ellipse, minimum width=2cm, minimum height=1cm, text centered, draw=black]
\tikzstyle{arrow} = [thick,<->,>=stealth]
\tikzstyle{line} = [thick,>=stealth]
\tikzstyle{darrow} = [thick,<->,>=stealth,dashed]
\tikzstyle{sarrow} = [thick,->,>=stealth]
\tikzstyle{larrow} = [line width=0.1mm,dashdotted,->,>=stealth]

\pgfkeys{/pgf/number format/.cd,1000 sep={\,}} 

\makeatletter
\def\grd@save@target#1{%
  \def\grd@target{#1}}
\def\grd@save@start#1{%
  \def\grd@start{#1}}
\tikzset{
  grid with coordinates/.style={
    to path={%
      \pgfextra{%
        \edef\grd@@target{(\tikztotarget)}%
        \tikz@scan@one@point\grd@save@target\grd@@target\relax
        \edef\grd@@start{(\tikztostart)}%
        \tikz@scan@one@point\grd@save@start\grd@@start\relax
        \draw[minor help lines] (\tikztostart) grid (\tikztotarget);
        \draw[major help lines] (\tikztostart) grid (\tikztotarget);
        \grd@start
        \pgfmathsetmacro{\grd@xa}{\the\pgf@x/1cm}
        \pgfmathsetmacro{\grd@ya}{\the\pgf@y/1cm}
        \grd@target
        \pgfmathsetmacro{\grd@xb}{\the\pgf@x/1cm}
        \pgfmathsetmacro{\grd@yb}{\the\pgf@y/1cm}
        \pgfmathsetmacro{\grd@xc}{\grd@xa + \pgfkeysvalueof{/tikz/grid with coordinates/major step x}}
        \pgfmathsetmacro{\grd@yc}{\grd@ya + \pgfkeysvalueof{/tikz/grid with coordinates/major step y}}
        \foreach \x in {\grd@xa,\grd@xc,...,\grd@xb}
        \node[anchor=north] at (\x,\grd@ya) {\pgfmathprintnumber{\x}};
        \foreach \y in {\grd@ya,\grd@yc,...,\grd@yb}
        \node[anchor=east] at (\grd@xa,\y) {\pgfmathprintnumber{\y}};
      }
    }
  },
  minor help lines/.style={
    help lines,
    gray,
    line cap =round,
    xstep=\pgfkeysvalueof{/tikz/grid with coordinates/minor step x},
    ystep=\pgfkeysvalueof{/tikz/grid with coordinates/minor step y}
  },
  major help lines/.style={
    help lines,
    line cap =round,
    line width=\pgfkeysvalueof{/tikz/grid with coordinates/major line width},
    xstep=\pgfkeysvalueof{/tikz/grid with coordinates/major step x},
    ystep=\pgfkeysvalueof{/tikz/grid with coordinates/major step y}
  },
  grid with coordinates/.cd,
  minor step x/.initial=.5,
  minor step y/.initial=.2,
  major step x/.initial=1,
  major step y/.initial=1,
  major line width/.initial=1pt,
}
\makeatother

\newlength\fheight
\newlength\fwidth

\makeatletter
\def\endthebibliography{%
	\def\@noitemerr{\@latex@warning{Empty `thebibliography' environment}}%
	\endlist
}
\makeatother

\def \sfwidth{0.9\linewidth}
\def \sfheight {0.5\linewidth}
\def \bpheight {0.4\columnwidth}

\definecolor{color0}{HTML}{FFD700}
\definecolor{color1}{HTML}{FFB14E}
\definecolor{color2}{HTML}{FA8775}
\definecolor{color3}{HTML}{EA5F94}
\definecolor{color4}{HTML}{CD34B5}
\definecolor{color5}{HTML}{9D02D7}
\definecolor{color6}{HTML}{0000FF}

\begin{document}

\title{Temporal Characterization of VR Traffic for Network Slicing Requirement Definition}

\author{Federico Chiariotti,~\IEEEmembership{Member,~IEEE,} Matteo Drago,~\IEEEmembership{Student Member,~IEEE,} \\
    Paolo Testolina,~\IEEEmembership{Student Member,~IEEE,} Mattia Lecci,~\IEEEmembership{Student Member,~IEEE,}\\
    Andrea Zanella,~\IEEEmembership{Senior Member, IEEE,} and Michele Zorzi,~\IEEEmembership{Fellow,~IEEE}
    \IEEEcompsocitemizethanks{\IEEEcompsocthanksitem{Federico Chiariotti (corresponding author, email: fchi@es.aau.dk) is with the Department of Electronic Systems, Aalborg University, 9220 Aalborg, Denmark. Matteo Drago (dragomat@dei.unipd.it), Paolo Testolina (testolina@dei.unipd.it), Mattia Lecci (leccimat@dei.unipd.it), Andrea Zanella (zanella@dei.unipd.it), and Michele Zorzi (zorzi@dei.unipd.it) are with the Department of Information Engineering, University of Padova, 35131 Padua, Italy.} \IEEEcompsocthanksitem{This work was partially supported by the National Institute of Standards and Technology (NIST) under award no. 60NANB21D127 and by the IntellIoT project under the H2020 framework grant no. 957218.
            The work of Mattia Lecci and Paolo Testolina was supported by Fondazione CaRiPaRo under grant ``Dottorati di Ricerca'' 2018 and 2019, respectively.}}}

\markboth{This paper has been submitted to IEEE Transactions on Mobile Computing. Copyright may change without notice.}%
{Chiariotti \MakeLowercase{\textit{et al.}}: Temporal Characterization of Virtual Reality Traffic for Network Slicing Requirement Definition}

\IEEEtitleabstractindextext{%
    \begin{abstract}
        Over the past few years, the concept of \gls{vr} has attracted increasing interest thanks to its extensive industrial and commercial applications. Currently, the 3D models of the virtual scenes are generally stored in the \gls{vr} visor itself, which operates as a standalone device. However, applications that entail multi-party interactions will likely require the scene to be processed by an external server and then streamed to the visors. 
        However, the stringent \glsfirst{qos} constraints imposed by \gls{vr}'s interactive nature require \glsfirst{ns} solutions, for which profiling the traffic generated by the \gls{vr} application is crucial. To this end, we collected more than 4 hours of traces in a real setup and analyzed their temporal correlation. More specifically, we focused on the \gls{cbr} encoding mode, which should generate more predictable traffic streams. From the collected data, we then distilled two prediction models for future frame size, which can be instrumental in the design of dynamic resource allocation algorithms. Our results show that even the state-of-the-art H.264 \gls{cbr} mode can have significant fluctuations, which can impact the \gls{ns} optimization.
        We then exploited the proposed models to dynamically determine the \glsfirst{sla} parameters in an \gls{ns} scenario, providing service with the required \gls{qos} while minimizing resource usage.
    \end{abstract}

    }

\IEEEdisplaynontitleabstractindextext

\maketitle

\glsresetall

\IEEEraisesectionheading{\section{Introduction}}
\label{sec:introduction}
\glsresetall


\IEEEPARstart{O}{ver} the past few years, the rapid technological development of \glspl{hmd} and the strong push towards the virtual world caused by the COVID-19 pandemic led to an explosion of the~\gls{xr} market, which includes technologies such as \gls{vr}, \gls{ar}, and \gls{mr}. Recent studies estimate hundreds of millions of users of these technologies in a time span of just 3 years~\cite{huaweiVrArWhitePaper}, requiring millions of new devices to be developed, produced, and shipped around the world for a business in the order of billions of dollars~\cite{huaweiArInsight}.

While the latest news on the \emph{metaverse} seem to indicate that the fastest growth will be in the entertainment and social media industries, \gls{xr} is expected to make an impact in a wide variety of scenarios~\cite{oculusVr,qualcommXr}. Interactive design, marketing, healthcare, and employee training are just a few of the proposed use cases, but industrial remote control in manufacturing and agriculture might have the largest impact, allowing human operators to remotely control machines in risky, hard to reach or unsafe environments, through a fully interactive virtual framework. 

A common characteristic of all these new applications is their interactive nature: users do not passively receive the information or stream a video, but need to manipulate the environment while maintaining an illusion of presence that requires the application to operate under very strict end-to-end delay constraints~\cite{3gpp.26.928,itu-t-f.743.10}. In particular, safety-critical and industrial applications will have even stricter constraints, as the consequences of network impairments can be significantly more serious. \emph{Cybersickness} is another important issue, as a delay over 20~ms between movements and visual and auditory feedback can cause disorientation and dizziness~\cite{huaweiVrArWhitePaper,kim2017vrSickness}.

In order to fulfill these stringent latency requirements over a wireless connection, the application and the network need to cooperate. The \emph{\gls{ns}} paradigm~\cite{barakabitze20205g} allows 5G and Beyond networks to reserve resources to a given stream, defining \gls{qos} targets.
Most works in this area, however, focus on relatively predictable applications.
In this setting, the need for predictability in the \gls{xr} traffic becomes extremely important, leading to a resurgence of \textit{quasi}-\gls{cbr} encoders, which are not used in non-interactive streaming due to their lower picture quality stability. While some efforts have been devoted by prominent standard bodies on this topic~\cite{3gpp.26.928, itu-t-f.743.10}, the current availability of traffic models for \gls{xr} is scarce. Furthermore, to the best of our knowledge, no detailed analysis of the temporal statistics of \textit{quasi}-\gls{cbr} video streams can be found in the literature, making existing slicing schemes rely on uncertain foundations. This makes the definition of a \gls{sla} for \gls{xr} traffic more complex: most \gls{ns} solutions assume that each application's demand in terms of required throughput and latency is known, but such a characterization can be difficult in case of streams with variable frame size, requiring significant overprovisioning.

However, even \gls{cbr} encoders are not perfect, and the interplay between the video content and the movements and actions of the users may cause significant fluctuations. In this work, we analyze the traffic from a real \gls{vr} application using the \textit{Periodic-Intra Refresh} mode of the H.264 codec, which results in relatively small differences in the frame sizes.
Modeling these imperfections, and consequently predicting the size of future frames in advance, can be extremely significant in the allocation of network resources, particularly if some critical \gls{qos} metrics have to be reached.
For example, this is the case for \textit{Cloud XR}, a new trend pursued by some major players in the telecommunications industry that moves the processing and rendering steps of the \gls{xr} content from the user to the Cloud, making the \gls{qos} requirements even more critical~\cite{nokiaCloudGaming,huawei2017cloudVr}. This increases the need for a dynamic \gls{sla} that can allow an \gls{ns} system to provide low-latency service to \gls{xr} applications without wasting too many resources: as we mentioned above, a static definition of the required throughput would require significant overprovisioning, while a temporal model of the \gls{xr} traffic stream could allow to predict the future needs of the application, tailoring the \gls{qos} requirements in the \gls{sla} to what will actually be needed.

Hence, in this paper we address the problem of providing a realistic stochastic characterization of a \gls{vr} traffic source, so as to allow for a dynamic provisioning of bandwidth resources for \gls{vr} users to satisfy the latency constraints. Our analysis can also be applied to the downlink part of generic \gls{xr} traffic.

Building upon our previous works~\cite{lecci21bursty,lecci2021open}, in which we collected more than 4 hours of live sessions and performed basic traffic characterization, in this paper we take the analysis one step further by modeling the size of \gls{vr} frames in the stream as a correlated time series that is then used to derive an adaptive and predictive \gls{sla}.
The contributions of this paper are the following:
\begin{itemize}
    \item We propose two parametric regression models to predict the size of future frames, and show that these models can be generalized to other traces and even different applications with limited regression performance loss;
    \item We analyze the residual error of these predictors, providing a full statistical model of future frame sizes;
    \item We show that the prediction can be successfully used for efficient resource allocation in an \gls{ns} scenario;
    \item We consider different \gls{ns} modes, including per-user or application-level slicing, and compare the performance of different schemes in terms of the trade-off between resource utilization and latency.
\end{itemize}
A partial version of this work was presented in~\cite{lecci22temporal}. 
This work significantly extends it by exploring the statistical analysis of frame sizes at a deeper level, including the characterization of the residual error of the predictors, and expanding on the \gls{sla} definition, including the use case with multiple \gls{vr} users.
All our traces, as well as the analysis and simulation code, are publicly available.\footnote{Code repository: {https://github.com/signetlabdei/vr-trace-analysis}}

The rest of the paper is structured as follows.
\cref{sec:soa} will discuss the current state of the art on \gls{xr} modeling, and our experimental setup is briefly presented in~\cref{sec:xr_streaming_architecture}.
Our analysis is reported in~\cref{sec:video_trace_analysis}, while~\cref{sec:network_slicing_use_case} illustrates how our analysis can be leveraged for a simple \gls{ns} use case by designing predictive resource allocation mechanisms and testing their performance in a simple simulation scenario.
Finally,~\cref{sec:conclusions} draws conclusions and presents some avenues for future work.

\section{State of the Art}
\label{sec:soa}

Despite a steady scientific interest in \gls{vr} since the 1990s~\cite{latta1994conceptual}, relatively little work has been done to characterize the details of this type of traffic.
We can distinguish four main areas of research: the reduction of the \gls{mtp} latency, the modeling and characterization of \gls{xr} traffic, the use of traffic models in \gls{ns}, and the scheduling and resource management of \gls{xr} data streams.

\subsection{Motion-To-Photon latency and VR sickness}

The \gls{mtp} latency is defined as the time difference between the beginning of a movement of the user's head and the instant when the image that corresponds to the user motion is shown on the HMD screen.
This phenomenon is one of the main factors causing sickness when experiencing \gls{xr} content, the main symptoms being discomfort, nausea, cold sweating, eye fatigue, and disorientation.

From a research point of view, a lot of effort has been devoted to avoiding such episodes in the first place, and the IEEE issued a dedicated standard in 2021~\cite{ieee2021vrstandard}, which addressed the content design, sickness assessment and measurement, and the network requirements that may influence the \gls{mtp} latency as the three main areas to consider in order to reduce or control bad user experience.

First of all, measuring the \gls{mtp} latency represents a challenge \textit{per se}. The architecture described in~\cite{minwoo2017photosensor}, which consists of a control PC, a head position model-based rotary platform, a pixel luminance change detector which converts the change in the display into a voltage value, and a digital oscilloscope to show such voltage information, is used as reference in~\cite{ieee2021vrstandard}. Specifically, after a movement of the rotary platform is generated by the control PC, the \gls{mtp} latency is measured as the time difference between the platform's movement and the corresponding change of the voltage value given by the oscilloscope.

To obtain a robust estimate of the \gls{mtp} latency, a precise head tracking algorithm is of the utmost importance. The authors of~\cite{blate2019implementation} presented a \gls{6dof}, optical head tracking instrument with a declared motion-to-pose latency (i.e., the time between a change in the users pose and the tracker actually detecting the movement) of about 28~$\mu$s, at a sample rate of 50~kHz. In their work, they also showed that the difference between the tracker's pose output and the user's true pose is proportional to pose velocity, tracker sampling rate, tracking latency, and noise.
Moreover, the authors of~\cite{stauffert2020simultaneous} showed that latency jitter artifacts already occur with a low system load by injecting artiﬁcial latency in a \gls{vr} simulation. Even though their hypothesis included the tracking algorithm of the speciﬁc \gls{hmd} as the possible cause of such jitter spikes, they did not prove it empirically.

Both rotational and translational \gls{mtp} latencies were estimated in~\cite{jingbo2017estimating} by calculating the phase shift between the captured signals of the physical motion of the \gls{hmd} and a motion-dependent gradient stimulus rendered on the display.
They were able to conclude that rapid head movements may elicit stronger disorientation to users in \gls{vr} environments than slower head movements do.
Even though the measurements were carried out with an Oculus Rift DK2, the proposed methodology is general and can be applied to other \glspl{hmd} as well.
The authors of~\cite{electronics7090171} also measured the \gls{mtp} latency with different workloads (determined by the complexity of the scene to render), finding that it can span from a minimum of 45~ms to a maximum of 155~ms. In general, the network requirements defined by the standard~\cite{ieee2021vrstandard} are way more stringent: approximately 5~ms for the wireless transmission and 20~ms in total for the \gls{mtp} latency, with a jitter strictly lower than 5~ms.

\subsection{XR Traffic Characterization}
\gls{xr} traffic modeling is closely related to 2D video content, and, even more so, to live, interactive applications such as video conferencing and gaming.
However, most of the work on the subject has considered the customary encoding schemes for pre-recorded video streaming, i.e., the \gls{vbr} encoding based on either the H.264 or the H.265 standard~\cite{tanwir2013vbrSurvey}.
\gls{vbr} can provide a stable visual quality, improving the user's \gls{qoe}, but is also subject to significant jitter due to the large frame size fluctuations.
Transmitting \gls{vbr} videos with low latency can then be a significant challenge even over channels with constant capacity~\cite{liew98vbrOverCbr}.
On the other hand, \gls{cbr} encoding sacrifices some visual quality stability to obtain an encoded video stream with a stable transmission rate~\cite{mohsenian99cbr}.
Although the higher predictability of the encoded output makes \gls{cbr} encoding attractive for interactive video and \gls{xr} content, it is still relatively unexplored in the relevant literature.

A topic related to \gls{xr} traffic is video game streaming, also called \textit{Cloud gaming}.
Games run on remote servers and the scenes are streamed directly to the users without the need for client-side computation.
The stringent requirements of gaming applications, especially in terms of latency, and the need to address them with optimized protocols and new transmission strategies, have led to an increased interest in their characterization.

The authors of~\cite{carrascosa2020stadia} carried out an extensive measurement campaign in Google Stadia, a popular Cloud gaming platform, giving an overview of its inner working.
They studied the distributions of downlink traffic, packet size and inter-packet time under multiple settings, including different resolutions, video codecs, and network conditions.
On the other hand, in~\cite{didomenico2021analysis,graff2021gaminganalysis} direct comparisons were made between different Cloud gaming platforms, mostly focusing only on the bit rate of the video stream, without including latencies or user experience.

A more comprehensive Cloud gaming testbed, including automated trace acquisition over Ethernet, WiFi, and LTE, was presented in~\cite{pulla2021measuring}.
Automating the acquisitions surely gives an advantage in terms of reproducibility and speed of the experiments, but the unpredictability of the users' actions in gaming scenarios (and, more importantly, in \gls{xr}) is the real challenge that the network has to face, limiting the usefulness of the results.
These works represent a good starting point for the collection and modeling of \gls{vr} traffic, as it is reasonable to assume that most of these Cloud gaming companies will start providing \gls{vr} services soon.
However, most works still focused on simple applications, such as interactive data visualization~\cite{hentschel2009vrSimulation}, and do not provide much insight on more complex scenarios.
There is an extensive literature on immersive video streaming~\cite{chiariotti2021vrSurvey}, but it has been mostly focused on passive applications in which the user is only a viewer, with different \gls{qoe} and encoding considerations. A recent document by the 3GPP~\cite{3gpp.38.838} also provided a simple model for \gls{xr} traffic, which however does not consider temporal or video content aspects, and is thus usable for general feasibility studies, but not for fine-grained optimization.

\subsection{Prediction-Based Slicing}

The use of traffic prediction in \gls{ns} is a concept that was first explored in~\cite{sciancalepore2017mobile}: as slicing requires precise \glspl{sla} to provide \gls{qos} to different services, but most practical applications are \gls{vbr}, characterizing the traffic and predicting future requests is a way to allocate resources in a foresighted manner, performing resource allocation on a short timescale and admission control on a longer one. If we consider wider networks with massive numbers of users, the daily, weekly, and seasonal cycles of network usage can also be exploited to allocate resources more effectively~\cite{marquez2019resource}.
This work, however, will focus on shorter-term predictions over a limited number of \gls{xr} flows. Other works such as~\cite{jiang2018multi} have also explored the possibility of exploiting longer-term trends for a rough resource allocation, while using a short-term scheduler for fine-grained optimization.

In particular, the use of \gls{arma} models has been explored in~\cite{tang2019arma} as a potential application-agnostic prediction method to perform resource allocation: as the orchestrator knows the state of the packet buffer for each slice, it can perform the moving average and allocate resources accordingly. However, \gls{arma} models require a certain number of past samples, and this approach cannot discriminate between different applications: consequently, the initial performance will be lower when compared to an application-aware model that takes knowledge of the traffic source into account. In order to capture the behavior of more complex traffic sources, it is also possible to replace the \gls{arma} model with a \gls{lstm} deep neural network~\cite{gutterman2019ran}, which can generalize to non-linear and longer-term patterns. Deep reinforcement learning~\cite{van2019real} is another alternative, as it can implicitly learn even complex application behaviors and take them into account when slicing.

Another recent idea is to combine \gls{ns} with video bit rate adaptation: if we consider \gls{qoe} as a flexible metric over which we can compromise in high traffic conditions, a cross-layer approach allows the orchestrator to dictate the video bit rate for the next few frames~\cite{arfaoui2020flexible}, limiting the demands of the interactive video flow to what the network is able to support. This approach is complementary to the prediction-based one, as these bit rate changes need to be relatively infrequent to avoid annoying the user, and such a system would operate over a longer timescale: while the prediction and allocation of resources is usually performed over tens or hundreds of milliseconds, video bit rate adaptation spans multiple seconds, and the two approaches can be integrated.

\subsection{XR Resource Management}

Although the management of \gls{xr} flows is a relatively new problem, a few works have already discussed efficient schemes for providing \gls{qos} to these applications. For example, in~\cite{chen18vrOverWireless,chen19correlation} game-theoretic approaches are proposed to tackle the optimization of multi-user \gls{vr} streaming over a small cell, with the help of machine learning.
The authors of~\cite{yang18mecVr} analyze the scheduling problem from the perspective of \gls{mec}, proposing scheduling strategies and analyzing communication, computing, and caching trade-offs.
While the models proposed for the network architectures considered in these works are extremely complex, there is no comparison with real-world \gls{vr} streaming.

To the best of our knowledge, our previous works, which proposed a simple architecture for collecting traffic traces from \gls{vr} games~\cite{lecci21bursty} and a simple generative model for the frame size~\cite{lecci2021open}, were the first to use real \gls{vr} traffic traces.
This paper extends our previous works by characterizing the temporal behavior of the \gls{vr} traces and drawing novel conclusions for \gls{ns} optimization.

\section{VR Streaming Architecture}
\label{sec:xr_streaming_architecture}

In this section, we describe the architecture of our \gls{vr} streaming acquisition and give some perspective on the full end-to-end setup.
To further understand what are the steps that most influence the \gls{vr} performance, it is useful to describe a common end-to-end \gls{vr} architecture.
First, we can start from the collection and processing of tracking information, delegated to the \gls{hmd}.
Then, this information is sent to a remote server to compose the viewport, i.e., what is actually shown to the user.
This process includes the rendering of the scene, the video encoding providing a more robust transmission towards the mobile device, and possibly some additional information, e.g., the direction in which the rendered frame is supposed to be displayed.
After receiving and decoding the video stream together with all the additional meta-information, the \gls{hmd} generates the images to display at the occurring screen refresh rate.
These steps need to be accomplished with minimal delay to guarantee adequate \gls{qoe}.

Our experimental setup consisted of a desktop computer equipped with an NVIDIA GeForce RTX~2080~Ti graphics card acting as the rendering server, and an iPhone~XS enclosed in a \gls{vr} cardboard acting as the \gls{hmd}.
VR applications were thus run on the rendering server and streamed to the headset using the \emph{RiftCat~2.0} application (on the server), and \emph{VRidge} 2.7.7 (on the phone).\footnote{\url{https://riftcat.com/vridge}}

The application uses hardware-accelerated H.264 encoding via \gls{nvenc} as long as a compatible graphics card is available to the system.
\emph{RiftCat}'s developers disclosed that Periodic Intra-Refresh is used, a setting provided by the encoder that allows each frame to be roughly the same size, making the stream almost \gls{cbr} and thus easier to handle from a network perspective.
It does so by replacing key frames with \emph{waves} of refreshed intra-coded blocks, i.e., blocks without any dependence on other frames, effectively spreading a key frame over multiple frames.
Image quality is balanced with resilience to packet loss by setting the \texttt{intraRefreshPeriod}, a parameter that determines the period after which an intra refresh happens again, and the \texttt{intraRefreshCnt} parameter, which sets the number of frames over which the intra refresh happens~\cite{nvencApi}.
If we consider a 30~\gls{fps} video, a value of 30 for the \texttt{intraRefreshPeriod} would ensure that the frame is fully recovered every second.
On the other hand, choosing small values of \texttt{intraRefreshCnt} leads to a quicker refresh but lower quality. 

Detailed information about the video encoder is fundamental for our work, since different encoders typically behave differently, especially when analyzing the temporal behavior of the encoded source.
Still, we believe that our work offers network researchers a peek into the intricacies of this topic, showing some key results on how a \gls{vr} traffic flow can be analyzed for resource provisioning.

Different freely available games and applications were used to acquire our dataset, including \emph{Minecraft}, \emph{Virus Popper}, and \emph{Google Earth VR}.
Further details on the acquisition setup and our traces can be found in~\cite{lecci2021open}. In the following, we will mostly concentrate on one trace acquired using the \emph{Virus Popper} application, but the methodology holds throughout the dataset, and can be easily replicated for any of the other traces.

\section{Video Trace Analysis}
\label{sec:video_trace_analysis}

By analyzing the acquired traces, we determined that the application used \gls{udp} over IPv4. It also used an additional application-layer protocol header of variable size, which we decoded to determine the types of the exchanged packets. More specifically, synchronization and acknowledgment packets were exchanged in both directions, while the \gls{ul} stream from the \gls{hmd} to the rendering server also contained frequent and relatively small head-tracking information packets. Naturally, the \gls{dl} stream also had regular video frame packet bursts.

\begin{figure*}[t!]
    \setlength\fheight{0.3\linewidth}
    \setlength\fwidth{0.9\linewidth}
    \centering
%
%

\definecolor{color0}{rgb}{1,0.498039215686275,0.0549019607843137}
\definecolor{color1}{rgb}{0.172549019607843,0.627450980392157,0.172549019607843}
\definecolor{color2}{rgb}{0.580392156862745,0.403921568627451,0.741176470588235}
\definecolor{color3}{rgb}{0.549019607843137,0.337254901960784,0.294117647058824}
\definecolor{color4}{rgb}{0.12156862745098,0.466666666666667,0.705882352941177}
\definecolor{color5}{rgb}{0.83921568627451,0.152941176470588,0.156862745098039}







\begin{tikzpicture}
\pgfplotsset{every tick label/.append style={font=\scriptsize}}

\begin{axis}[%
width=0,
height=0,
at={(0,0)},
scale only axis,
xmin=0,
xmax=0,
xtick={},
ymin=0,
ymax=0,
ytick={},
axis background/.style={fill=white},
legend style={legend cell align=center, align=center, draw=white!15!black, font=\scriptsize, at={(0, 0)}, anchor=center, /tikz/every even column/.append style={column sep=2em}},
legend columns=4,
]

\addplot [thick, color1, mark=triangle*, mark size=3, mark options={solid,rotate=180}, only marks]
table {%
0 0
};
\addlegendentry{Video Frame (DL)}

\addplot [thick, color4, mark=triangle*, mark size=3, mark options={solid,rotate=180}, only marks]
table {%
0 0
};
\addlegendentry{Frame Feedback (DL)}

\addplot [thick, color0, mark=triangle*, mark size=3, mark options={solid}, only marks]
table {%
0 0
};
\addlegendentry{Frame Feedback (UL)}

\addplot [thick, color5, mark=triangle*, mark size=3, mark options={solid}, only marks]
table {%
0 0
};
\addlegendentry{Head Tracking (UL)}

\end{axis}
\end{tikzpicture}%

    \input{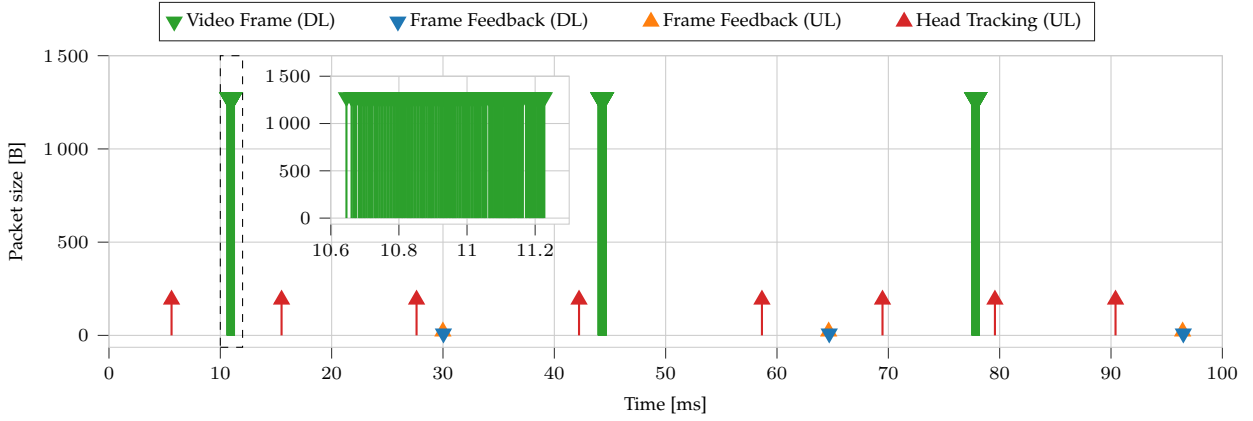}

    \caption{Portion of traffic trace from \emph{Virus Popper} (50~Mb/s, 30~FPS). In this trace, each video frame burst consists in about 130--140 individual fragments.}
    \label{fig:stem}
\end{figure*}

\begin{figure}[t!]
    \setlength\fheight{0.5\columnwidth}
    \setlength\fwidth{0.9\columnwidth}
    \centering
    \input{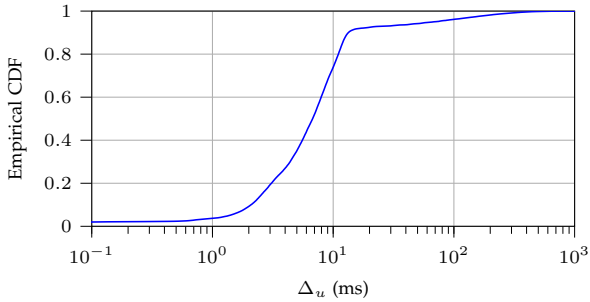}
    \caption{Head tracking packet inter-arrival time.}
    \label{fig:headTrackingIpi}
\end{figure}

\cref{fig:stem} is a visual representation of a short period of bidirectional \gls{vr} streaming, showing the main data streams in both \gls{dl} and \gls{ul}.
As the figure clearly shows, most of the traffic is concentrated in \gls{dl} and consists of packet bursts encoding video frames.
Video frame fragments were consistently found to be 1320~B long in all acquired traces, with a data size (the UDP payload) of 1278~B.

First, we considered the head tracking packets in the \gls{ul}, which were all 192~B long. The distribution of the inter-packet interval $\Delta_u$ is shown in Fig.~\ref{fig:headTrackingIpi}: tracking packets are relatively frequent, with a median interval of about 7~ms, but the distribution has a long tail. This suggests that head-tracking packets are usually sent at regular intervals, but some are transmitted adaptively if there were significant headset movements that can affect the video rendering on the \gls{hmd}. As we did not manage to decode the content of the tracking packets, a deeper analysis of their relation to head movements is left as future work.

By decoding the application protocol, we managed to identify frame boundaries and sort out the video data frames from metadata and control information.
We can then consider the size of individual frames in a video trace.
We note that non-video packets have a low impact on the total streaming data rate.
Considering this, as well as the strong dependence of metadata on the application setup, we decided to focus mostly on the video frame data, discarding all other packets from our analysis.
Our results can then be applied to any \gls{vr} application using the same encoder.

The encoder uses the H.264 \emph{Periodic Intra-Refresh} compression scheme to reduce the variation between frame sizes, so we do not expect a multimodal distribution, as would be the case for a classical keyframe-based encoding.
As we mentioned above, encoding \gls{vr} traffic as \gls{cbr} offers a significant advantage for the network optimization, because frames of constant size make it possible for \gls{ns} schemes to provide a guaranteed latency without wasting resources.

\begin{figure}[t!]
    \setlength\fheight{0.5\columnwidth}
    \setlength\fwidth{0.9\columnwidth}
    \centering
    \input{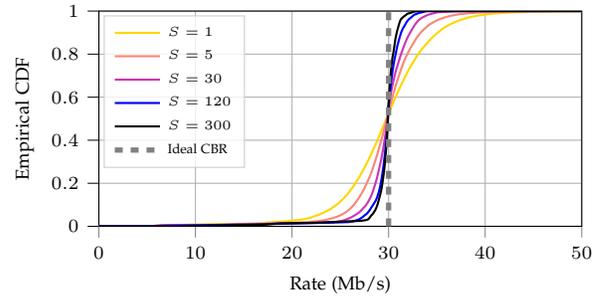}
    \caption{Rate distribution for different MA window sizes $S$ [number of frames].}
    \label{fig:rate_dist}
\end{figure}

However, \gls{cbr} encoding is not perfect, and frames may still have variable size, although the average rate almost perfectly matches the required one.
We can use a simple \gls{ma} filter with a rectangular window $S$ to examine the behavior of the traffic on longer timescales, which is useful if resource allocation is performed at a slower pace.
Naturally, allocating resources every $S$ frames leads to a larger jitter between frames, but it can also improve the resource allocation efficiency, as size fluctuations tend to average out over multiple frames.

In order to measure this effect, we consider the \emph{Virus Popper} trace, with a required rate $R=30$~Mb/s and a $\varphi=60$~\gls{fps} refresh rate.
We only measure the video traffic, without packet headers and redundancy added by the application, which results in an average rate of 29.76~Mb/s.
Fig.~\ref{fig:rate_dist} shows the empirical \gls{cdf} of the rate, considering different values of the \gls{ma} window sizes.
If we consider each frame individually, there is a significant variation in the rate, which gradually reduces as we increase the number of frames over which the rate is measured.

\begin{figure}[t!]
    \setlength\fheight{0.5\columnwidth}
    \setlength\fwidth{0.9\columnwidth}
    \centering
    \input{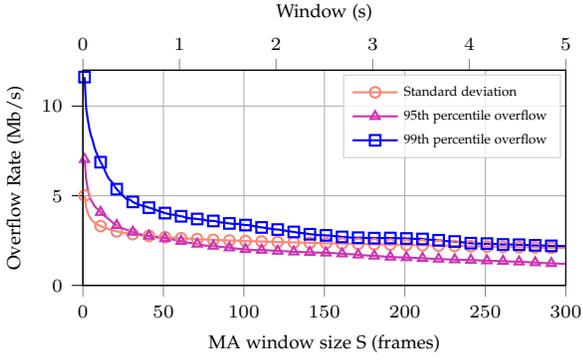}
    \caption{Overflow rate for a target CBR of 30 Mb/s.}
    \label{fig:rate_deviation}
\end{figure}

However, providing reliable service will require a significant overhead even if we relax the slicing time: Fig.~\ref{fig:rate_deviation} shows the overflow rate, i.e., the difference between the actual rate and the expected 30~Mb/s \gls{cbr} rate, as a function of the \gls{ma} window sizes.
The plot shows the standard deviation, as well as the 95\textsuperscript{th} and 99\textsuperscript{th} percentile overflow rates.
If our aim is to provide 99\% reliability, we need to overprovision by more than 8~Mb/s (i.e., almost 30\% of the \gls{cbr} rate) even if we consider a timescale of 100~ms for resource allocation, i.e., 6~frames.
Even averaging over periods of multiple seconds leads to worst-case rates almost 4~Mb/s higher than the target, probably because of highly dynamic content in the video.
Interestingly, the overflow standard deviation is approximately constant if the \gls{ma} window is longer than 50 frames, while the higher percentiles of the overflow continue to decay: this suggests that higher throughput periods tend to be shorter and more frequent, while there are longer periods of time with a bit rate below the average. Fig.~\ref{fig:rate_dist} also hints at a skew in the distribution, as the left tail of the frame size empirical \gls{cdf} is much longer.

\begin{figure}[t!]
    \setlength\fheight{0.5\columnwidth}
    \setlength\fwidth{0.9\columnwidth}
    \centering
\begin{tikzpicture}

\pgfplotsset{every tick label/.append style={font=\scriptsize}}

\begin{axis}[
width=\fwidth,
height=\fheight,
legend cell align={left},
legend style={font=\tiny, at={(0.99,0.98)}, anchor=north east, legend cell align=left, align=left,fill opacity=0.8, draw opacity=1, text opacity=1, draw=white!80!black},
tick align=outside,
tick pos=left,
x grid style={white!69.0196078431373!black},
xlabel={Lag (frames)},
xmin=0, xmax=50,
xmajorgrids,
y grid style={white!69.0196078431373!black},
ylabel={Autocorrelation},
ymin=-0.5, ymax=1,
ymajorgrids,
ytick style={color=black}
]

\addplot [thick, color2,mark=o]
table {%
0 1
1 0.517836309350073
2 0.413433745231653
3 0.325157432676376
4 0.334787919362688
5 0.36519833379723
6 0.356017247272571
7 0.328556538997123
8 0.304253705535198
9 0.266788043173347
10 0.255763516076102
11 0.2548017444313
12 0.26646456353622
13 0.255464803188973
14 0.236750594796484
15 0.220307147882125
16 0.223062177290464
17 0.229831506934101
18 0.222898388658022
19 0.223897545031599
20 0.230686767907829
21 0.234647400834151
22 0.230260534822164
23 0.226602491583914
24 0.217071994617214
25 0.227452603733461
26 0.213386068697139
27 0.215787871258549
28 0.221592055961187
29 0.224567490702634
30 0.218419964178907
31 0.210120189862408
32 0.211239130223141
33 0.212972097197209
34 0.210054525646951
35 0.202210908472013
36 0.202675305478668
37 0.212528836200849
38 0.203010795535743
39 0.195270971362609
40 0.192884017775452
41 0.194220521732189
42 0.194832131188118
43 0.184805020749906
44 0.190071612671639
45 0.191038955662475
46 0.197135402897756
47 0.193320784862826
48 0.186847058531481
49 0.180483883732332
50 0.183084442475377
};
\addlegendentry{Autocorr. of $F(t)$}
\addplot [thick, color6,mark=triangle]
table {%
0 1
1 -0.391773918934805
2 -0.0167388583966961
3 -0.101523244932691
4 -0.021531004237653
5 0.0410247990022447
6 0.0189691616296346
7 -0.00329634684125523
8 0.0136480089590587
9 -0.0274076688622721
10 -0.0104279178233368
11 -0.013082646070798
12 0.0234735500234375
13 0.0079971119581818
14 -0.00235404027385207
15 -0.0198941317471495
16 -0.00415761941443526
17 0.014192500934362
18 -0.00822239184843067
19 -0.00599209756683808
20 0.0029341319357817
21 0.00863783139347187
22 -0.000749942477685081
23 0.00607456771605123
24 -0.0206171505779058
25 0.0253087640063166
26 -0.0170456500865272
27 -0.00352288647866168
28 0.00291153497114848
29 0.00944091159789572
30 0.00223142835361686
31 -0.00973955903802881
32 -0.000639128580590911
33 0.00481533889499011
34 0.00508919076522155
35 -0.00859679376871189
36 -0.00970065563295453
37 0.0200334177820948
38 -0.0018446008745478
39 -0.0055280800472064
40 -0.00385233849658026
41 0.000753202118417935
42 0.0109999169994948
43 -0.0158194972698275
44 0.00444128984054233
45 -0.00528620020506456
46 0.0102407456333972
47 0.00274593261622197
48 -0.000119883283715285
49 -0.00925807406492241
50 0.00821833954332399
};
\addlegendentry{Autocorr. of $\Delta F(t)$}

\addplot [thick, gray, dashed, forget plot]
table {%
0 0.05
1 0.05
2 0.05
3 0.05
4 0.05
5 0.05
6 0.05
7 0.05
8 0.05
9 0.05
10 0.05
11 0.05
12 0.05
13 0.05
14 0.05
15 0.05
16 0.05
17 0.05
18 0.05
19 0.05
20 0.05
21 0.05
22 0.05
23 0.05
24 0.05
25 0.05
26 0.05
27 0.05
28 0.05
29 0.05
30 0.05
31 0.05
32 0.05
33 0.05
34 0.05
35 0.05
36 0.05
37 0.05
38 0.05
39 0.05
40 0.05
41 0.05
42 0.05
43 0.05
44 0.05
45 0.05
46 0.05
47 0.05
48 0.05
49 0.05
50 0.05
};
\addplot [thick, gray, dashed, forget plot]
table {%
0 -0.05
1 -0.05
2 -0.05
3 -0.05
4 -0.05
5 -0.05
6 -0.05
7 -0.05
8 -0.05
9 -0.05
10 -0.05
11 -0.05
12 -0.05
13 -0.05
14 -0.05
15 -0.05
16 -0.05
17 -0.05
18 -0.05
19 -0.05
20 -0.05
21 -0.05
22 -0.05
23 -0.05
24 -0.05
25 -0.05
26 -0.05
27 -0.05
28 -0.05
29 -0.05
30 -0.05
31 -0.05
32 -0.05
33 -0.05
34 -0.05
35 -0.05
36 -0.05
37 -0.05
38 -0.05
39 -0.05
40 -0.05
41 -0.05
42 -0.05
43 -0.05
44 -0.05
45 -0.05
46 -0.05
47 -0.05
48 -0.05
49 -0.05
50 -0.05
};

\end{axis}

\begin{axis}[
width=\fwidth,
height=\fheight,
tick align=outside,
tick pos=right,
xlabel={Lag (ms)},
xmin=0, xmax=833.33333333,
ymin=0, ymax=1,
ymajorticks = false,
]
    
\end{axis}

\end{tikzpicture}
    \caption{Video frame size autocorrelation for \emph{Virus Popper} (30~Mb/s, 60~\gls{fps}).}
    \label{fig:autocorr}
\end{figure}
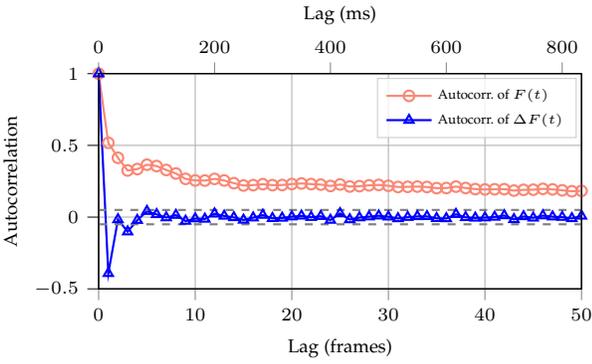

\begin{figure}[t!]
    \setlength\fheight{0.5\columnwidth}
    \setlength\fwidth{0.8\columnwidth}
    \centering
\begin{tikzpicture}

\begin{axis}[
width=\fwidth,
height=\fheight,
colormap={mymap}{[1pt]
 rgb(0pt)=(0.699611712320639,0.155783736277259,0.200466931944463);
  rgb(1pt)=(0.741098023510142,0.285713892997613,0.323031245695321);
  rgb(2pt)=(0.782584334699646,0.415644049717968,0.445595559446178);
  rgb(3pt)=(0.827030095357373,0.554842847640379,0.576903067825654);
  rgb(4pt)=(0.868516406546877,0.684773004360733,0.699467381576512);
  rgb(5pt)=(0.91000271773638,0.814703161081087,0.822031695327369);
  rgb(6pt)=(0.944495395711727,0.947045513158855,0.95014114060491);
  rgb(7pt)=(0.816034334821363,0.854298101102617,0.883181238459061);
  rgb(8pt)=(0.690058891103904,0.762308988051814,0.814457809630203);
  rgb(9pt)=(0.555835055238277,0.66429677831316,0.741234632485169);
  rgb(10pt)=(0.429121474155285,0.571768666553075,0.672108527315517);
  rgb(11pt)=(0.303146030437825,0.479779553502272,0.603385098486659);
  rgb(12pt)=(0.177170586720365,0.387790440451469,0.534661669657801)
  rgb(13pt)=(0.177170586720365,0.387790440451469,0.534661669657801)
},
colorbar sampled,
colormap access=piecewise constant,
colorbar right,
point meta max=1,
point meta min=-1,
colorbar style={ytick={-1,-0.75,-0.5,-0.25,0,0.25,0.5,0.75,1},
samples=14,
yticklabels={
  \ensuremath{-}1,
  \ensuremath{-}0.75,
  \ensuremath{-}0.50,
  \ensuremath{-}0.25,
  0,
  0.25,
  0.50,
  0.75,
  1
},ylabel={}},
tick align=outside,
tick pos=left,
x grid style={white!69.0196078431373!black},
xlabel={Lag (frames)},
xmin=0, xmax=61,
xtick={0.5,5.5,10.5,15.5,20.5,25.5,30.5,35.5,40.5,45.5,50.5,55.5,60.5},
xticklabels={0,5,10,15,20,25,30,35,40,45,50,55,60},
y dir=reverse,
y grid style={white!69.0196078431373!black},
ylabel={Time (s)},
ymin=0, ymax=566,
ytick style={color=black},
ytick={0.5,60.5,120.5,180.5,240.5,300.5,360.5,420.5,480.5,540.5},
yticklabels={
  0,60,120,180,240,300,360,420,480,540}
]
\addplot graphics [includegraphics cmd=\pgfimage,xmin=0, xmax=61, ymin=566, ymax=0] {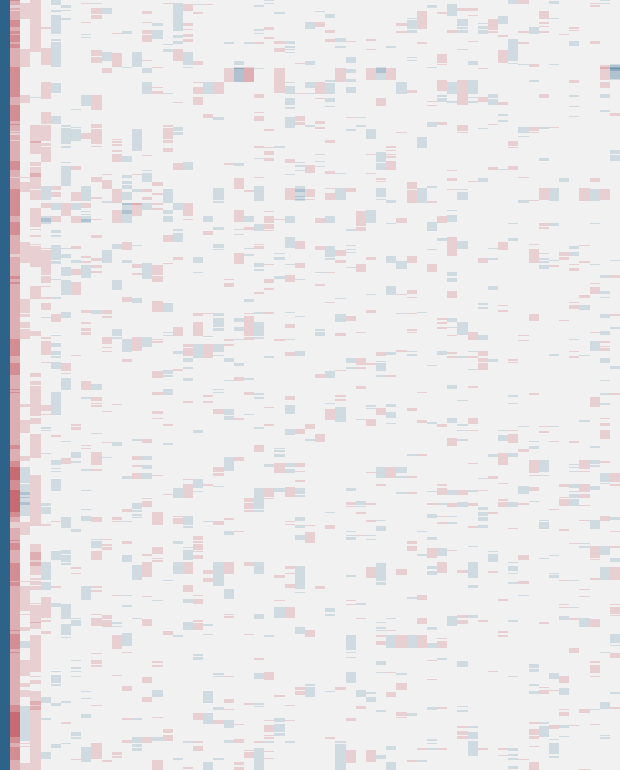};
\end{axis}

\begin{axis}[
  width=\fwidth,
  height=\fheight,
  tick align=outside,
  tick pos=right,
  xlabel={Lag (ms)},
  xmin=0, xmax=1016.667,
  ymin=0, ymax=1,
  ymajorticks = false,
  ]

  \end{axis}

\end{tikzpicture}
    \caption{Rolling windowed $\Delta F$ autocorrelation for \emph{Virus Popper} (30~Mb/s, 60~\gls{fps}). The windows were 600 frames (10 s) long, with a time shift of 60 frames (1 s).}
    \label{fig:autocorr_rolling}
\end{figure}

Finally, we can analyze the autocorrelation of the frame size sequence $F(t)$, to identify patterns in how the sequence changes.
Fig.~\ref{fig:autocorr} shows the autocorrelation of $F(t)$ and $\Delta F(t)=F(t)-F(t-1)$.
While $F(t)$ has a strong long-term autocorrelation, due to the constant component, the $\Delta F(t)$ sequence has a strong negative autocorrelation between one frame and the next, while almost all longer time differences fall within the $\pm$0.05 range.
This means that the encoder tends to balance out fluctuations between one frame and the next, such that a frame that is bigger than the previous one tends to be followed by a smaller one again.
We can check that this holds throughout the whole video by computing a rolling window autocorrelation, shown in Fig.~\ref{fig:autocorr_rolling} for $\Delta F(t)$.
In this case, the plot clearly shows that there are no strong long-term correlations in any part of the video.
The frame difference sequence has a noticeable and consistent autocorrelation only with lags 1, 3, and 5, confirming the result from Fig.~\ref{fig:autocorr}.

\subsection{Frame Size Prediction}

\begin{figure*}[t!]
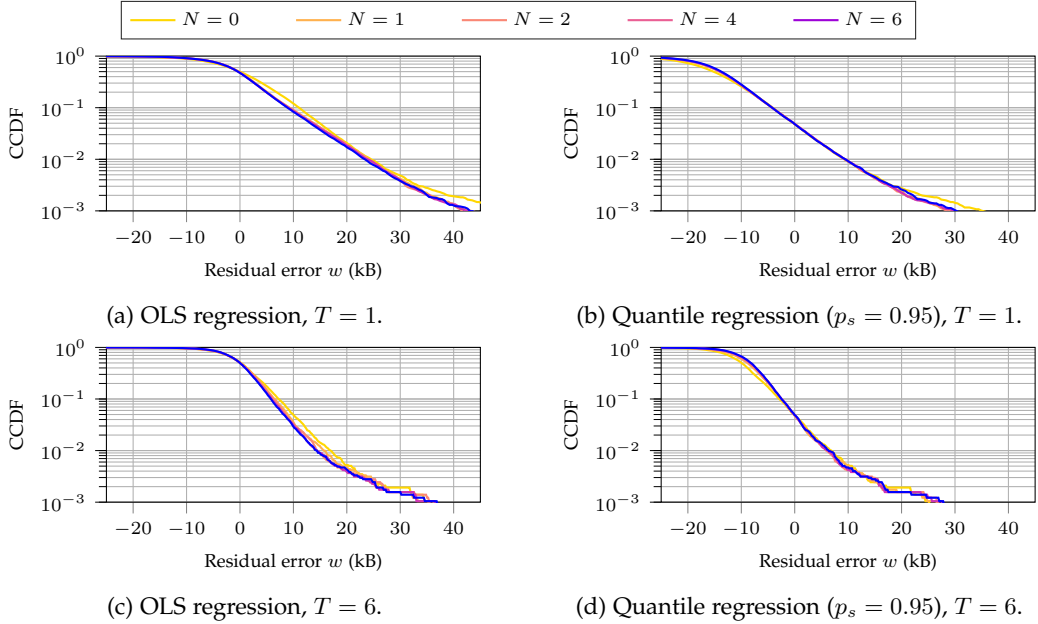

    \centering
    \begin{subfigure}[b]{0.9\linewidth}
        \centering
\begin{tikzpicture}

\begin{axis}[
    width=0,
    height=0,
    at={(0,0)},
    scale only axis,
    xmin=0,
    xmax=0,
    xtick={},
    ymin=0,
    ymax=0,
    ytick={},
    axis background/.style={fill=white},
    legend style={legend cell align=center, align=center, draw=white!15!black, font=\scriptsize, at={(0, 0)}, anchor=center, /tikz/every even column/.append style={column sep=2em}},
    legend columns=10,
]
\addplot [thick, color0]
table {%
0 1
};
\addlegendentry{$N=0$}
\addplot [thick, color1]
table {%
0 1
};
\addlegendentry{$N=1$}
\addplot [thick, color2]
table {%
0 1
};
\addlegendentry{$N=2$}
\addplot [thick, color3]
table {%
0 1
};
\addlegendentry{$N=4$}
\addplot [thick, color5]
table {%
0 1
};
\addlegendentry{$N=6$}
\end{axis}

\end{tikzpicture}
    \end{subfigure}
    \begin{subfigure}[b]{.4\linewidth}
        \centering
        \input{./img/ccdf_Linear_1_frames.tex}
        \caption{OLS regression, $T=1$.}
        \label{fig:o1_ccdf}
    \end{subfigure}
    \begin{subfigure}[b]{.4\linewidth}
        \centering
        \input{./img/ccdf_Quantile_1_frames.tex}
        \caption{Quantile regression ($p_s=0.95$), $T=1$.}
        \label{fig:q1_ccdf}
    \end{subfigure}
    \begin{subfigure}[b]{.4\linewidth}
        \centering
        \input{./img/ccdf_Linear_6_frames.tex}
        \caption{OLS regression, $T=6$.}
        \label{fig:o6_ccdf}
    \end{subfigure}
    \begin{subfigure}[b]{.4\linewidth}
        \centering
        \input{./img/ccdf_Quantile_6_frames.tex}
        \caption{Quantile regression ($p_s=0.95$), $T=6$.}
        \label{fig:q6_ccdf}
    \end{subfigure}
    \caption{Complementary CDF of the error $w$ with $\tau=1$ and different values of $N$ and $T$.}
    \label{fig:ccdf}
\end{figure*}

Let us consider the average size of future frames in the time interval $[t,t+T)$, given by
\begin{equation}\label{eq:model}
    \overline{F_T}(t) = \frac{1}{T} \sum_{i=0}^{T-1} F(t+i).
\end{equation}
We denote by $\hat{F}_T(t,\tau)$ an estimate of $\overline{F_T}(t+\tau)$, $\tau>0$, i.e., considering a look-ahead of $\tau$ frames.
We focus on linear predictors based on the last $N\geq 0$ samples, so that
\begin{equation}
    \hat{F}_T(t,\tau) =\theta_0 + \sum_{j=1}^{N} \theta_j F(t-j+1),
\end{equation}
where $\boldsymbol{\theta}=[\theta_0,\ldots,\theta_N]$ is a weight vector, which determines the accuracy of the estimate. If $N=0$, the estimate is just given by the parameter $\theta_0$, and does not consider any past frames.
The difference between actual and estimated value is captured by the error term
$
    w(t,\tau,T) = \overline{F_T}(t+\tau) - \hat{F}_T(t,\tau),
$
which will be denoted just as $w$ in the following, for ease of writing.
We can then consider two regression methods to determine the value of the parameter vector $\bm{\theta}$:
\begin{itemize}
    \item \emph{\gls{ols} linear regression}: least squares regression was independently developed by Gauss and Legendre in the 19\textsuperscript{th} century~\cite{stigler1981gauss}, and is the most classic form of regression. In this case, the objective is to minimize the $\ell^2$ norm of the sequence $w$. \gls{ols} regression can be useful in determining the average behavior of the underlying stochastic process, giving easily interpretable results on the quality of the prediction and the dynamics of the frame size over time;
    \item \emph{Quantile regression}~\cite{koenker1978regression}: this technique estimates $\hat F_T(t,\tau) $ so that the probability that it is higher than the real value, is not larger than $p_s$. This has obvious implications for network resource provisioning: as we are interested in providing enough resources to send a frame within the required latency with probability $p_s$, estimating the corresponding quantile might be the best way to get the required quality.
\end{itemize}
We also used \emph{Robust linear regression}~\cite{yu2017robust} to verify that the \gls{ols} prediction was not too sensitive to outliers. We considered a robust method using Huber's T norm instead of the $\ell^2$ norm: the two norms have the same quadratic behavior if the error is smaller than a threshold $\delta$, but Huber's T increases linearly for larger values. Setting the threshold to $\delta=\frac{\mathbb{E}\left[|F|\right]}{4}$, we found that the results matched exactly those of the \gls{ols} model, suggesting that outliers are not playing a relevant role in this case and thus letting us discard this model.

In this section, we will show results for both the \gls{ols} and the quantile regression models. As we stated above, while the results from \gls{ols} are more immediate, quantile regression is useful when focusing on scheduling network resources for a \gls{vr} stream, which requires a model of the tail of the frame size distribution to provide latency guarantees.

\begin{figure*}[t!]
    \centering
    \begin{subfigure}[b]{\linewidth}
        \centering
\begin{tikzpicture}

\begin{axis}[
    width=0,
    height=0,
    at={(0,0)},
    scale only axis,
    xmin=0,
    xmax=0,
    xtick={},
    ymin=0,
    ymax=0,
    ytick={},
    axis background/.style={fill=white},
    legend style={legend cell align=center, align=center, draw=white!15!black, font=\scriptsize, at={(0, 0)}, anchor=center, /tikz/every even column/.append style={column sep=2em}},
    legend columns=10,
]
\addplot [thick, color0]
table {%
0 1
};
\addlegendentry{$N=0$}
\addplot [thick, color1]
table {%
0 1
};
\addlegendentry{$N=1$}
\addplot [thick, color2]
table {%
0 1
};
\addlegendentry{$N=2$}
\addplot [thick, color3]
table {%
0 1
};
\addlegendentry{$N=4$}
\addplot [thick, color5]
table {%
0 1
};
\addlegendentry{$N=6$}
\end{axis}

\end{tikzpicture}
    \end{subfigure}
    \\
    \begin{subfigure}[b]{.4\linewidth}
        \centering
\begin{tikzpicture}

\begin{axis}[
width=\sfwidth,
height=\sfheight,
legend style={at={(0.98,0.98)}, anchor=north east, fill opacity=0.8, legend columns=4, draw opacity=1, text opacity=1,
draw=white!80!black, /tikz/every even column/.append style={column sep=0.5em}, font=\tiny},
tick align=outside,
tick pos=left,
x grid style={white!69.0196078431373!black},
xlabel={Lag (frames)},
xmin=-1, xmax=21,
xmajorgrids,
y grid style={white!69.0196078431373!black},
ylabel={Autocorrelation},
ymin=-0.153089121643665, ymax=1.05490900579256,
ymajorgrids,
ytick style={color=black}
]
\addplot [thick, color0]
table {%
0 1
1 0.467627024563451
2 0.386678461671627
3 0.289915198879574
4 0.275341597285425
5 0.270901758454847
6 0.248515400368811
7 0.221993838594262
8 0.174933283065541
9 0.13787191459132
10 0.108290797337813
11 0.0907287081702729
12 0.103473392129106
13 0.0753888838219881
14 0.0669010114992973
15 0.0489823718882972
16 0.0515623035432974
17 0.0498877306298421
18 0.0485075634479317
19 0.0468378237299123
20 0.0463075565134515
};
\addplot [thick, color1]
table {%
0 1
1 -0.0981801158511094
2 0.150727876249507
3 0.0569536333138058
4 0.0945702071523353
5 0.109681763589611
6 0.093295851119966
7 0.0933436899686838
8 0.0579502860841136
9 0.0458793165918818
10 0.0324117660134167
11 0.0151054264356603
12 0.0621405907530121
13 0.0158851546791096
14 0.0300677691084331
15 0.00567374053361614
16 0.021370106517629
17 0.018034155019033
18 0.0179125819417568
19 0.0164247763242153
20 0.0159557097716162
};
\addplot [thick, color2]
table {%
0 1
1 -0.0123542659112769
2 -0.0459725029274762
3 -0.00838855384686502
4 0.0372228213004587
5 0.0798062695607738
6 0.0740336677715648
7 0.0707189477427878
8 0.0356486651473089
9 0.0191725256450366
10 0.00334588427917296
11 0.000745876565648191
12 0.0494437424866286
13 0.013536039878729
14 0.0134577681478106
15 -0.00411508966893306
16 0.0103456153901497
17 0.0128301491257589
18 0.0113924843772721
19 0.00962478497343389
20 0.0123376736821672
};
\addplot [thick, color4]
table {%
0 1
1 -0.00767622651644151
2 -0.00911913219876571
3 -0.0243293418515051
4 -0.050330227437615
5 0.0521471524304755
6 0.0544249216265398
7 0.0613449419629661
8 0.0217760181946534
9 0.00515678478885353
10 -0.0101410214492291
11 -0.0130000163398945
12 0.0373749631472741
13 0.00339960442326097
14 0.00820049224484625
15 -0.00946794358294854
16 0.00236726905841527
17 0.00650022868050762
18 0.00751618487076797
19 0.00754877690226283
20 0.00746307120497348
};
\addplot [thick, color6]
table {%
0 1
1 -0.000759254633870117
2 -0.000853197018060813
3 0.000448584907611879
4 0.00333930346158908
5 -0.000139496948214762
6 -0.00470407465790527
7 0.0334446501055545
8 0.00905558940152654
9 -0.00197130860538306
10 -0.0194313680077943
11 -0.0257626732512656
12 0.0227010693515244
13 -0.00971851568058316
14 -0.00269143150622811
15 -0.0170476495799799
16 -0.000632848764992775
17 0.00150593700937746
18 0.000554235200813429
19 0.000489356756473908
20 0.00419233471715153
};
\addplot [thick, black, dashed, forget plot]
table {%
0 0.05
1 0.05
2 0.05
3 0.05
4 0.05
5 0.05
6 0.05
7 0.05
8 0.05
9 0.05
10 0.05
11 0.05
12 0.05
13 0.05
14 0.05
15 0.05
16 0.05
17 0.05
18 0.05
19 0.05
20 0.05
};
\addplot [thick, black, dashed, forget plot]
table {%
0 -0.05
1 -0.05
2 -0.05
3 -0.05
4 -0.05
5 -0.05
6 -0.05
7 -0.05
8 -0.05
9 -0.05
10 -0.05
11 -0.05
12 -0.05
13 -0.05
14 -0.05
15 -0.05
16 -0.05
17 -0.05
18 -0.05
19 -0.05
20 -0.05
};



\end{axis}

\begin{axis}[
width=\sfwidth,
height=\sfheight,
tick align=outside,
tick pos=right,
xlabel={Lag (ms)},
xmin=-16.6666, xmax=350,
ymin=0, ymax=1,
ymajorticks = false,
]
    
\end{axis}

\end{tikzpicture}
        \caption{OLS regression.}
        \label{fig:ols_aut}
    \end{subfigure}
    \begin{subfigure}[b]{.4\linewidth}
        \centering
\begin{tikzpicture}

\begin{axis}[
width=\sfwidth,
height=\sfheight,
legend style={at={(0.98,0.98)}, anchor=north east, fill opacity=0.8, legend columns=4, draw opacity=1, text opacity=1,
draw=white!80!black, /tikz/every even column/.append style={column sep=0.5em}, font=\tiny},
tick align=outside,
tick pos=left,
x grid style={white!69.0196078431373!black},
xlabel={Lag (frames)},
xmin=-1, xmax=21,
xmajorgrids,
y grid style={white!69.0196078431373!black},
ylabel={Autocorrelation},
ymin=-0.1025, ymax=1.0525,
ymajorgrids,
ytick style={color=black}
]
\addplot [thick, color0]
table {%
0 1
1 0.467627024563451
2 0.386678461671627
3 0.289915198879574
4 0.275341597285425
5 0.270901758454847
6 0.248515400368811
7 0.221993838594262
8 0.174933283065541
9 0.13787191459132
10 0.108290797337813
11 0.0907287081702729
12 0.103473392129106
13 0.0753888838219881
14 0.0669010114992973
15 0.0489823718882972
16 0.0515623035432974
17 0.0498877306298421
18 0.0485075634479317
19 0.0468378237299123
20 0.0463075565134515
};
\addplot [thick, color1]
table {%
0 1
1 0.11185971871572
2 0.238317020566131
3 0.143432481945961
4 0.161675722595726
5 0.169530017466258
6 0.150913296625593
7 0.141098357843621
8 0.101372039057096
9 0.0800297389797738
10 0.0605786526595123
11 0.0431762006619358
12 0.0774847018057679
13 0.0379718738943583
14 0.0437391953861639
15 0.021748034004522
16 0.0325757555063692
17 0.0298597477040893
18 0.0292694013075912
19 0.0277163440116886
20 0.027224178574118
};
\addplot [thick, color2]
table {%
0 1
1 0.130716678672006
2 0.0521744616693844
3 0.0671832300271736
4 0.0969660351526297
5 0.129566599805081
6 0.120363435609184
7 0.110219527695984
8 0.0718538070756958
9 0.049168733548824
10 0.0292502901586177
11 0.0241436002676665
12 0.0630366816871386
13 0.0304631840422267
14 0.02644213288973
15 0.00953966755844923
16 0.0205611009803969
17 0.0225482128839955
18 0.0209776184070289
19 0.0192094579069057
20 0.0213786643399552
};
\addplot [thick, color4]
table {%
0 1
1 0.12783148979721
2 0.0601079926304476
3 0.00924952628325991
4 -0.016922614179137
5 0.0846953382532696
6 0.0887697814940289
7 0.0901403405638285
8 0.0468926982594005
9 0.0233695400868274
10 0.00512705693340227
11 0.0022784062159266
12 0.0451842755832299
13 0.0156887399056038
14 0.0163361194148314
15 -0.000954710119289624
16 0.00829132812481772
17 0.0126714535180997
18 0.0141355704391201
19 0.0144651878416944
20 0.0138534701947519
};
\addplot [thick, color6]
table {%
0 1
1 0.121686208292072
2 0.0553768086562304
3 0.0243079745939452
4 0.0174688233218054
5 -0.00218919932627444
6 -0.000863150983135271
7 0.0439628528749094
8 0.0201273257392078
9 0.00443485933169177
10 -0.0160839034781266
11 -0.0224038854611669
12 0.0199508097986374
13 -0.00683705769588252
14 -0.00216797375647116
15 -0.0147706793723008
16 9.17449698461189e-05
17 0.00266124490489729
18 0.00226018784294586
19 0.00284471017179433
20 0.00673860883475947
};
\addplot [thick, black, dashed, forget plot]
table {%
0 0.05
1 0.05
2 0.05
3 0.05
4 0.05
5 0.05
6 0.05
7 0.05
8 0.05
9 0.05
10 0.05
11 0.05
12 0.05
13 0.05
14 0.05
15 0.05
16 0.05
17 0.05
18 0.05
19 0.05
20 0.05
};
\addplot [thick, black, dashed, forget plot]
table {%
0 -0.05
1 -0.05
2 -0.05
3 -0.05
4 -0.05
5 -0.05
6 -0.05
7 -0.05
8 -0.05
9 -0.05
10 -0.05
11 -0.05
12 -0.05
13 -0.05
14 -0.05
15 -0.05
16 -0.05
17 -0.05
18 -0.05
19 -0.05
20 -0.05
};



\end{axis}

\begin{axis}[
width=\sfwidth,
height=\sfheight,
tick align=outside,
tick pos=right,
xlabel={Lag (ms)},
xmin=-16.6666, xmax=350,
ymin=0, ymax=1,
ymajorticks = false,
]
    
\end{axis}

\end{tikzpicture}
        \caption{Quantile regression ($p_s=0.95$).}
        \label{fig:quant_aut}
    \end{subfigure}
    \caption{Autocorrelation of the residual error $w$ for next-frame prediction ($T=1$, $\tau=1$) for different values of $N$.}
    \label{fig:autocorr_pred}
\end{figure*}
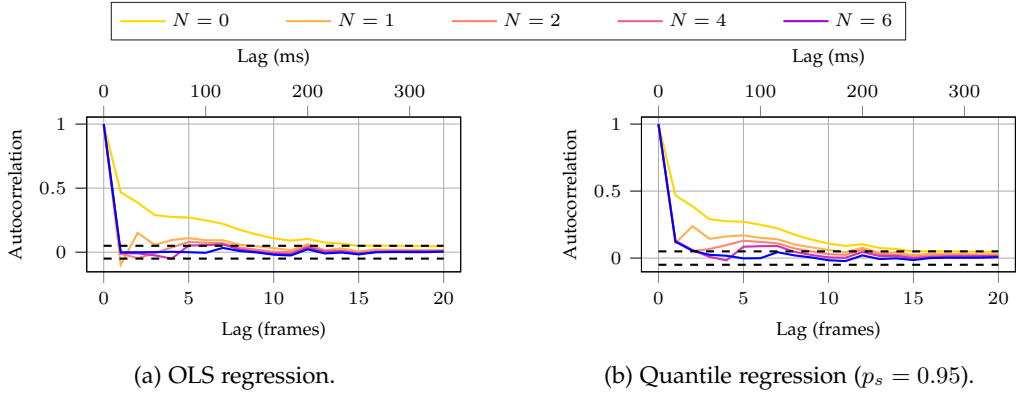

We can now examine the results of the regression analysis for the \emph{Virus Popper} trace, considering a rate of 30~Mb/s and 60~\gls{fps}.
We focus on this video trace as the standard example in the paper, but other traces, even at different bit rates and frame rates, exhibit a similar behavior.
Fig.~\ref{fig:ccdf} shows the complementary \gls{cdf} of the residual error $w$, considering $\tau=1$ and two different values of the averaging interval $T$.
The first thing we can notice is that the error distribution has a slightly different shape for the \gls{ols} and quantile regression models, indicating that the difference in the two models is not simply a shift in the value of the intercept $\theta_0$, but instead the two predictions are inherently different.
We can also notice that there is some benefit from having a longer memory, although increasing $N$ yields diminishing returns in terms of increased accuracy.
Finally, we can confirm that the reliable transmission of this \gls{vr} content will require significant overprovisioning, even when using prediction: for $T=1$ the 95th percentile error of the \gls{ols} prediction is approximately 15~kB higher than the mean with any of the models, i.e., about 25\% of the average frame size (which is 62.5~kB for this trace).
In fact, this is close to the difference between the average predictions of the \gls{ols} and the quantile models.

This difference is about halved for $T=6$, due to the fact that computing the average over multiple frames allows errors to compensate and cancel each other out.
However, provisioning over multiple frames means that only the average amount of resources will be assigned to the stream, which will cause larger frames to have a higher latency, thus causing additional queuing delay to subsequent frames.
Since the frame cannot be properly shown on screen until it is fully received, this translates to a higher jitter and reduces the \gls{qoe} perceived by the user, making a lower value of $T$ preferable.

Another fundamental component in evaluating the quality of a predictor is the autocorrelation of the residual error $w$: if the autocorrelation between subsequent samples of the residual error is high, the model did not capture some effect, usually due to an insufficient memory, i.e., too low a value of $N$.
Fig.~\ref{fig:autocorr_pred} shows the autocorrelation of $w$ for different values of $N$: it is easy to see that models with $N<4$, and particularly with $N=0$ and $N=1$, do not have enough memory to capture the frame size dynamics.
This is more evident in quantile regression, which shows a higher autocorrelation for these models.

\begin{figure}[t!]
    \centering
    \begin{tikzpicture} 
    \begin{axis}[
    name=hm1,
    width=\bpheight,
    height=\bpheight,
    xlabel=$N$,
    title={OLS},
    ylabel=$\tau$,
    mesh/cols=11,
    mesh/rows=11,
    xmin=-0.5,
    xmax=10.5,
    ymin=-0.5,
    ymax=9.5,
    point meta min=8.4,
    point meta max=10.6,
colormap={mymap}{[1pt]
 rgb(0pt)=(0.946403,0.937159,0.458592);
  rgb(1pt)=(0.962517,0.851476,0.285546);
  rgb(2pt)=(0.981173,0.759135,0.156863);
  rgb(3pt)=(0.987945,0.667748,0.058329);
  rgb(4pt)=(0.981895,0.579392,0.02625);
  rgb(5pt)=(0.961293,0.488716,0.084289);
  rgb(6pt)=(0.929644,0.411479,0.145367);
  rgb(7pt)=(0.886302,0.342586,0.202968);
  rgb(8pt)=(0.832299,0.283913,0.257383);
  rgb(9pt)=(0.769556,0.236077,0.307485);
  rgb(10pt)=(0.694627,0.195021,0.354388);
  rgb(11pt)=(0.621685,0.164184,0.388781);
  rgb(12pt)=(0.547157,0.136929,0.413511);
  rgb(13pt)=(0.472328,0.110547,0.428334);
  rgb(14pt)=(0.397674,0.083257,0.433183);
  rgb(15pt)=(0.316282,0.05349,0.425116);
  rgb(16pt)=(0.238273,0.036621,0.396353);
  rgb(17pt)=(0.15585,0.044559,0.325338);
  rgb(18pt)=(0.081962,0.043328,0.215289);
  rgb(19pt)=(0.025793,0.01933,0.10593);
},
    ytick={1,3,5,7,9},
    yticklabels={2,4,6,8,10}
]
    \addplot[matrix plot*, point meta=explicit] file [meta=index 2] {./img/imgdata/Linear_std.csv};
\end{axis}

  \begin{axis}[
  name=hm2,
    at=(hm1.right of south east), anchor=left of south west,
    width=\bpheight,
    height=\bpheight,
    title={Quantile},
    xlabel=$N$,
    ylabel=$\tau$,
    mesh/cols=11,
    mesh/rows=11,
    xmin=-0.5,
    xmax=10.5,
    ymin=-0.5,
    ymax=9.5,
    point meta min=8.4,
    point meta max=10.6,
colormap={mymap}{[1pt]
 rgb(0pt)=(0.946403,0.937159,0.458592);
  rgb(1pt)=(0.962517,0.851476,0.285546);
  rgb(2pt)=(0.981173,0.759135,0.156863);
  rgb(3pt)=(0.987945,0.667748,0.058329);
  rgb(4pt)=(0.981895,0.579392,0.02625);
  rgb(5pt)=(0.961293,0.488716,0.084289);
  rgb(6pt)=(0.929644,0.411479,0.145367);
  rgb(7pt)=(0.886302,0.342586,0.202968);
  rgb(8pt)=(0.832299,0.283913,0.257383);
  rgb(9pt)=(0.769556,0.236077,0.307485);
  rgb(10pt)=(0.694627,0.195021,0.354388);
  rgb(11pt)=(0.621685,0.164184,0.388781);
  rgb(12pt)=(0.547157,0.136929,0.413511);
  rgb(13pt)=(0.472328,0.110547,0.428334);
  rgb(14pt)=(0.397674,0.083257,0.433183);
  rgb(15pt)=(0.316282,0.05349,0.425116);
  rgb(16pt)=(0.238273,0.036621,0.396353);
  rgb(17pt)=(0.15585,0.044559,0.325338);
  rgb(18pt)=(0.081962,0.043328,0.215289);
  rgb(19pt)=(0.025793,0.01933,0.10593);
},
    colorbar sampled,
    colormap access=piecewise constant,
    colorbar right,
    colorbar style={samples=21,
                    ylabel={$\sigma_w$ (kB)},
                    ytick={7, 7.5, ..., 15}},
    ytick={1,3,5,7,9},
    yticklabels={2,4,6,8,10}
]
    \addplot[matrix plot*, point meta=explicit] file [meta=index 2] {./img/imgdata/Quantile_std.csv};
\end{axis}
\end{tikzpicture}
    %
    \caption{Heatmap of the residual error standard deviation (measured in kB) as a function of $N$ and $\tau$, with $T=1$.}
    \label{fig:heat}
\end{figure}
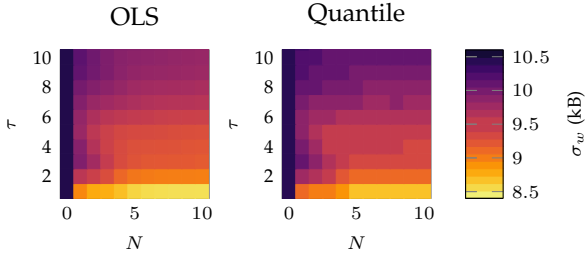

Finally, we can examine the effect of $N$ and $\tau$ on the quality of the prediction by looking at Fig.~\ref{fig:heat}, which shows the standard deviation of the residual error $w$ as a function of these two parameters with $T=1$.
The figure clearly shows that increasing the memory of the model improves the prediction, but gives diminishing returns, as the difference between $N=6$ and $N=10$ is minimal.
Furthermore, we see an expected increase of the error if $\tau$ increases, but this is not monotonic for $N<3$: this might be due to the autocorrelation we observed in the $w$ sequence, as $N<3$ is not sufficient to fully represent the state of the stochastic process, resulting in suboptimal predictions.

\subsection{Residual Error Characterization}

\begin{figure}
    \setlength\fheight{0.5\columnwidth}
    \setlength\fwidth{0.9\columnwidth}
    \centering
    \input{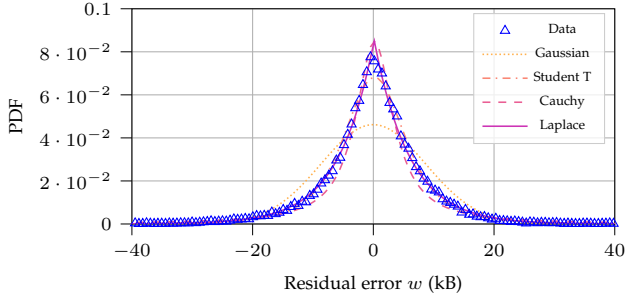}
    \caption{Residual error after \gls{ols} prediction for \emph{Virus Popper} (30~Mb/s, 60~\gls{fps}), with $T=1$, $\tau=1$, and $N=6$.}
    \label{fig:residue_dist}
\end{figure}

We can then analyze the residual error $w$ in more detail: as Fig.~\ref{fig:residue_dist} shows, we attempted to fit the residual error on the frame size to various common bilateral distributions, and the maximum likelihood fit was given by the Laplace$(\mu,b)$ distribution, whose \gls{pdf} is given by:
\begin{equation}
    p_w(x;\mu,b)=\frac{1}{2b}e^{-\frac{|x-\mu|}{b}},\label{eq:laplace_pdf}
\end{equation}
where $\mu$ is the location parameter and $b$ is the shape parameter. The same result held for all other traces in the dataset, leading us to infer that this distribution depends on some inherent property of the encoder and the way it generates frames, instead of specific features in the video content.

If we consider the residual error of the \gls{ols} regression method, the best estimate of the parameter $\mu$ is $\hat{\mu}=0$, as having a non zero-mean residual error would imply a bias in the \gls{ols} estimator. The maximum likelihood estimator of the shape parameter $b$ is then given in~\cite{norton1984double} by:
\begin{equation}
    \hat{b}=\frac{1}{N}\sum_{i=1}^N |x_i-\hat{\mu}|.
\end{equation}
The instantaneous value $\hat{b}_T(t,\tau)$ can then be determined from the model.
As we are considering a regression model, in which previous frame sizes affect the distribution of future frames, we can simply perform an \gls{ols} regression on the absolute value of the residual error $|w|$ to find $\hat{b}_T(t,\tau)$. We can then represent the future frame size as a value $\hat{F}_T(t,\tau)$ given by the prediction plus a noise term $w$, whose distribution is Laplace$(\hat{F}_T(t,\tau),\hat{b}_T(t,\tau))$. 
Interestingly, if we adopt this model, we have the complete distribution of the frame size, making it extremely easy to derive the quantile values for any desired point and considerably reducing the computational impact with respect to multiple quantile regressions. The quantile function $P^{-1}(p_s|T,\tau,t)$ is given by:
\begin{equation}
    P^{-1}(p_s|T,\tau,t)=\hat{F}_T(t,\tau)+\hat{b}_T(t,\tau)\log(\min(2p_s,2-2p_s)).\label{eq:laplace_icdf}
\end{equation}

\begin{figure*}[t!]
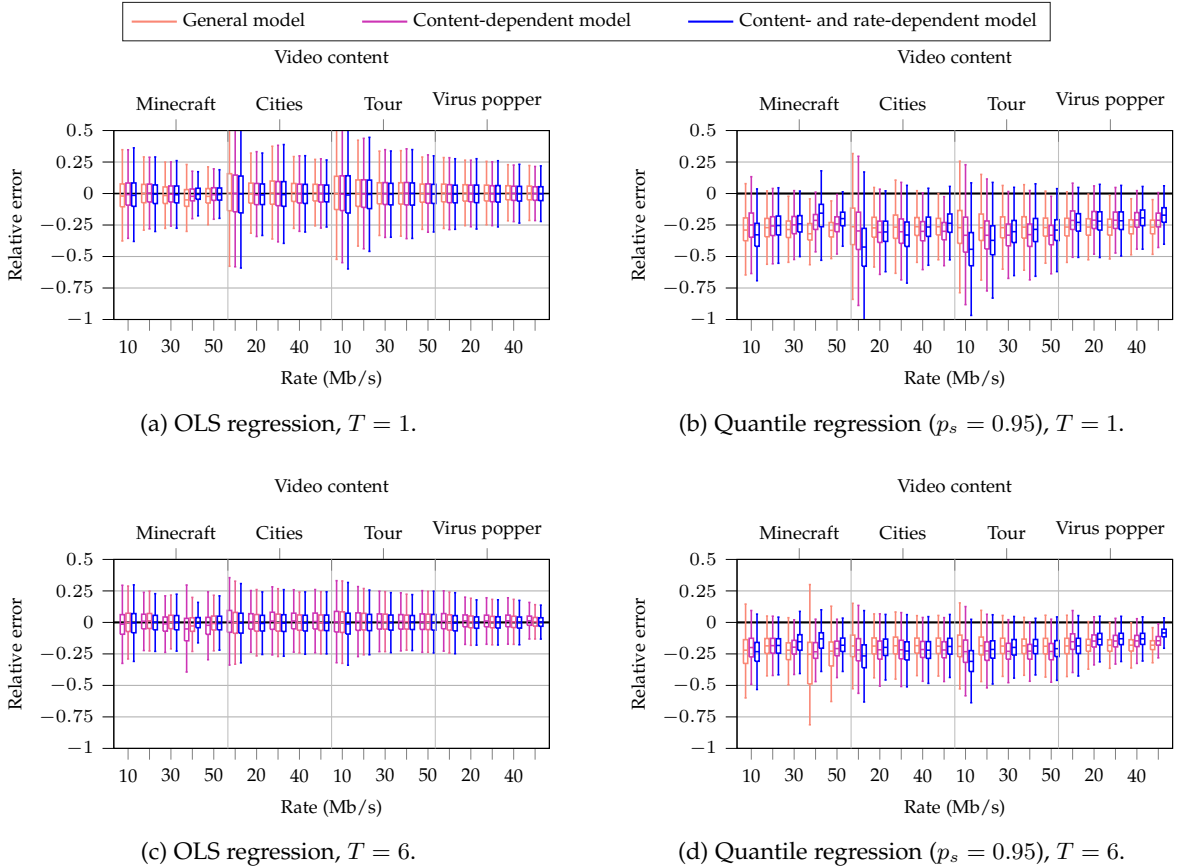

    \centering
    \begin{subfigure}[b]{\linewidth}
        \centering
\begin{tikzpicture}

\begin{axis}[
    width=0,
    height=0,
    at={(0,0)},
    scale only axis,
    xmin=0,
    xmax=0,
    xtick={},
    ymin=0,
    ymax=0,
    ytick={},
    axis background/.style={fill=white},
    legend style={legend cell align=center, align=center, draw=white!15!black, font=\scriptsize, at={(0, 0)}, anchor=center, /tikz/every even column/.append style={column sep=2em}},
    legend columns=10,
]
\addplot [thick, color2]
table {%
0 1
};
\addlegendentry{General model}
\addplot [thick, color4]
table {%
0 1
};
\addlegendentry{Content-dependent model}
\addplot [thick, color6]
table {%
0 1
};
\addlegendentry{Content- and rate-dependent model}

\end{axis}

\end{tikzpicture}
    \end{subfigure}
    \\
    \begin{subfigure}[b]{.45\linewidth}
        \centering
        \input{./img/gen_boxplot_Linear_1.tex}
        \caption{OLS regression, $T=1$.}
        \label{fig:ols_box1}
    \end{subfigure}
    \begin{subfigure}[b]{.45\linewidth}
        \centering
        \input{./img/gen_boxplot_Quantile_1.tex}
        \caption{Quantile regression ($p_s=0.95$), $T=1$.}
        \label{fig:quant_box1}
    \end{subfigure}
    \\\bigskip
    \begin{subfigure}[b]{.45\linewidth}
        \centering
        \input{./img/gen_boxplot_Linear_6.tex}
        \caption{OLS regression, $T=6$.}
        \label{fig:ols_box6}
    \end{subfigure}
    \begin{subfigure}[b]{.45\linewidth}
        \centering
        \input{./img/gen_boxplot_Quantile_6.tex}
        \caption{Quantile regression ($p_s=0.95$), $T=6$.}
        \label{fig:quant_box6}
    \end{subfigure}
    \caption{Boxplot of the relative residual error $\frac{\varphi w}{R}$ for different levels of generalization with $N=6$ and $\tau=1$. The traces are grouped by video content, and each group of boxplots shows the error at different bit rates for that video content.}
    \label{fig:box_gen}
\end{figure*}

\subsection{Model Generalization}\label{ssec:general}

In the above, we studied how well regression models can predict future frame sizes $\hat{F}_T(t,\tau)$, but we always found the parameter vector $\bm{\theta}$ based on the same video trace. In the following, we study how prediction models perform when the regression is performed over multiple traces, with different bit rates and types of content. This has significant advantages, as finding a predictor for each specific video content requires acquiring traces for each content and quality level, while generalizing the predictor would allow for simpler deployment.

We consider $N=6$ and $\tau=1$, as we determined that $N=6$ is sufficient to capture the dynamics of the model.
In order to directly compare traces with different bit rates $R$ and frame rates $\varphi$, we normalize the video traces by the expected frame size $\varphi^{-1}R$, obtaining a normalized parameter vector $\tilde{\bm{\theta}}$, which, given the linearity of our models, can be converted back to the original parameter vector in~\eqref{eq:model} as $\bm{\theta}=\frac{R\tilde{\bm{\theta}}}{\varphi}$. By normalizing our frame sizes, we can train and use our models on multiple traces with different values of $R$ and $\varphi$.
We then consider three generalized models:
\begin{enumerate}
    \item A \gls{gm}, which computes $\bm{\theta}$ using the whole dataset, with different frame rates, bit rates, and video content types;
    \item A \gls{cm}, which computes $\bm{\theta}$ using a single type of content (e.g., the \emph{Virus Popper} game), but with different bit rates and frame rates;
    \item A \gls{crm}, which derives the parameter vector on a per-content, frame rate, and bit rate basis, i.e., a single trace.
\end{enumerate}

Given that different values of $R$ and $\varphi$ can have different scales of errors which can be difficult to compare directly, in \cref{fig:box_gen} we show the error normalized to the expected frame size $R/\varphi$.
As the figure shows, the model can generalize quite well: the performance of \gls{cm} is almost always similar to that obtained by \gls{crm}, making generalization across different bit rates and frame rates possible for the same video content.
On the other hand, \gls{gm} performs slightly worse, and has a large error in the Minecraft trace with $R=40$~Mb/s: it is possible that this trace involves different dynamics in the content or head movements, leading to sharp differences even with other traces with the same type of content.
On the other hand, \gls{gm} has similar performance to \gls{cm} and \gls{crm} with the \gls{ols} predictor, but shows a less consistent behavior for the quantile regressor.
For example, the \textit{Minecraft} trace with $R=40$~Mb/s shows very different performance between the three models and different values of $T$. Furthermore, the \emph{Virus Popper} trace seems to have a smaller tail, as \gls{gm} is more conservative than the models based only on that video content.

As we can see, using the quantile model leads to a prediction between 25\% and 40\% higher than the average, skewing the error distribution.
We should also further highlight that averaging over multiple frames can also significantly reduce the error across almost all traces.

\section{Predictive Network Slicing}
\label{sec:network_slicing_use_case}

In this section, we consider an \gls{ns} use case for the models we developed in Sec.~\ref{sec:video_trace_analysis}. We assume that a number $M$ of \gls{vr} clients are assigned to a high-priority slice, with the objective of allowing each frame to be delivered before the generation of the next one, i.e., maintaining a latency below $1/60^{\rm th}$ of a second. Provisioning the time and frequency resources for \gls{vr} is a critical component of Beyond 5G networks, and guaranteeing limited latency while reducing the impact on other users is an important application of our model. Each client $m$ then has a different bit rate $R_m$ and, potentially, a different application, but we assume that the clients all share the same frame rate $\varphi$. Furthermore, each client $m$ has a different spectral efficiency $\eta_m$, depending on its connection's \gls{snr}: users closer to the base station will have a stronger signal, and consequently, a higher transmission efficiency.

With a small loss of generality, we assume that clients are synchronized, i.e., frames are generated at the same time. The orchestration can be adapted relatively easily to the more general case, but the notation would be much more cumbersome, and we maintain this simplifying assumption for the sake of readability. We can then assume that the network slicing orchestrator is equipped with the general frame size distribution model from Sec.~\ref{ssec:general}, and can estimate the frame size distribution for arbitrary values of $T$ and $\tau$ for each client $m$. We consider an orchestrator that can make decisions on the resource allocation only at times $t=kS$, $k\in \mathbb{Z}$, i.e., every $S$~frames or, conversely, every $\Delta t = \frac{S}{\varphi}$~ms. In the following, we consider queued bits from earlier frames in the slicing as well.
At time $t=kS$, we consider that the previous slice might have been unable to send all the data in time, leaving in the queue $q_m(t)$ bits that have to be sent in the following frame intervals.

\subsection{Motion-To-Photon Latency}

\begin{figure}[t]
    \centering
    \resizebox{0.9\columnwidth}{!}{\begin{tikzpicture}[auto]

\node[name=hmd_t] at (0,5) {HMD};
\node[name=hmd_b] at (0,-0.5) {};
\node[name=bs_t] at (3,5) {BS};
\node[name=cl_t] at (6,5) {Cloud};
\node[name=bs_b] at (3,-0.5) {};
\node[name=cl_b] at (6,-0.5) {};

\draw[-] (hmd_t) to (hmd_b);
\draw[-] (bs_t) to (bs_b);
\draw[-] (cl_t) to (cl_b);

\draw[densely dotted] (0,4.5) to (7,4.5);
\draw[densely dotted] (3,4) to (7,4);
\draw[densely dotted] (6,3) to (7,3);
\draw[densely dotted] (6,2) to (7,2);
\draw[densely dotted] (6,2.5) to (7,2.5);
\draw[densely dotted] (3,1.5) to (7,1.5);
\draw[densely dotted] (0,-0.2) to (7,-0.2);

\draw[|-|] (0,4) to (0,3);
\draw[-|] (0,3) to (0,2);
\draw[-|] (0,2) to (0,1);
\draw[-|] (0,1) to (0,0);

\draw[|-|] (6,4) to (6,2.5);
\draw[|-] (6,1) to (6,2.5);

\draw[dashed,|-|] (6.25,4) to node[midway,right] {$\varphi^{-1}$} (6.25,2.5);
\draw[|-|] (7,4.5) to node[midway,right] {$T(k)$} (7,-0.2);

\draw[dashed,|-|] (-0.25,4.5) to node[midway,left] {$\tau_m$} (-0.25,4);
\draw[dashed,|-|] (0.25,3) to node[midway,right] {$\Delta_u$} (0.25,2);
\draw[dashed,->] (0,4) to (3,4);
\draw[dashed,->] (3,4) to node[midway,above] {$\tau_p$} (6,3);
\draw[dashed,|-|] (5.75,2) to node[midway,left] {$\tau_r$} (5.75,2.5);
\draw[dashed,|-|] (5.75,3) to node[midway,left] {$\tau_f$} (5.75,2.5);
\draw[dashed,->] (6,2) to node[midway,above] {$\tau_p$} (3,1);
\draw[dashed,->] (3,1) to node[midway,above] {$\frac{ F(k)}{C(k)}$} (0,-0.2);

\end{tikzpicture}}
    \caption{Schematic of the components of the \gls{mtp} latency.}
    \label{fig:mtp_lat}
\end{figure}
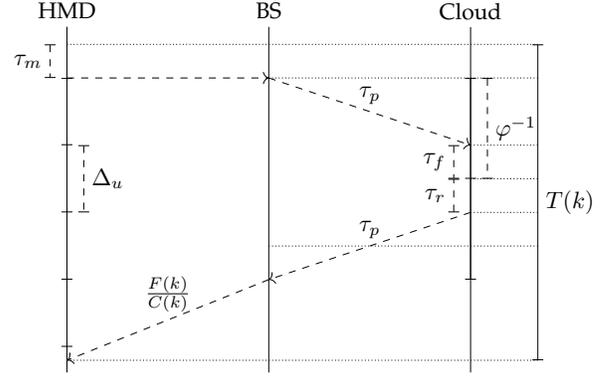

We can now analyze the \gls{mtp} latency by dividing it into 6 components, which are shown in Fig.~\ref{fig:mtp_lat}:
\begin{enumerate}
    \item The movement of the user needs to be recorded and transmitted. For simplicity, we can assume head tracking packets to be transmitted at a constant interval $\Delta_u$; in this case, the time between the motion and its transmission is $\tau_m\sim\mathcal{U}(0,\Delta_u)$.
    \item The head tracking packet needs to be transmitted to the Cloud \gls{vr} server. Considering that the uplink traffic is very light, as the packet is small, we can assume that it only incurs a constant propagation delay $\tau_p$. We can also assume that $\Delta_u$ includes the uplink transmission time to the \gls{bs}, simplifying the model. In general, the transmission time from the \gls{hmd} to the \gls{bs} should be extremely low.
    \item The head tracking data is received by the Cloud server, which then needs to produce a frame. The frame generation delay is $\tau_f\sim\mathcal{U}(0,\varphi^{-1})$.
    \item The server needs to generate, render, and encode the frame. We denote this delay as $\tau_r$, and assume that it is constant across short periods of time.
    \item The frame is transmitted to the \gls{bs} through a series of fiber optic links. As the capacity of fiber optic links is much higher than the \gls{ran}'s, we can assume this to take only the propagation time $\tau_p$.
    \item The frame is transmitted from the \gls{bs} to the \gls{hmd}. This component depends on both the frame size and the downlink bandwidth allocated to its slice by the orchestrator.
\end{enumerate}

If we set a maximum allowed \gls{mtp} latency $T_{\max}$, we can then derive a condition on the minimum bandwidth $B(k)$ to be assigned to the $n$-th customer in the $k$-th interval of time:
\begin{equation}
    B(k)\geq\frac{F(k)}{\eta\left(T_{\max}-\tau_m-2\tau_p-\tau_f-\tau_r\right)}.
\end{equation}
where $\eta$ is the spectral efficiency, known to the \gls{bs}.
However, $\tau_m$ and $\tau_f$ are random variables, so we can set a stricter condition that guarantees that the latency requirement is met in the worst case by substituting their maximum values, i.e., $\Delta_u$ and $\varphi^{-1}$, respectively.
We hence obtain
\begin{equation}
    B(k)\geq\frac{F(k)}{\eta\left(T_{\max}-\Delta_u-2\tau_p-\varphi^{-1}-\tau_r\right)}.
\end{equation}
For the sake of readability, we denote the maximum time allowed for the \gls{ran} transmission to fulfill the \gls{mtp} latency requirement as $T_{\text{tx}}$, i.e.,
\begin{equation}
    T_{\text{tx}}=T_{\max}-\Delta_u-2\tau_p-\varphi^{-1}-\tau_r.
\end{equation}
Finally, to ensure the stability of the queue at the \gls{bs}, the average allocated bit rate, $\eta\mathbb{E}[B(k)]$, must be larger than the mean offered traffic, i.e.,
\begin{equation}
    \eta\mathbb{E}[B(k)]>\varphi\mathbb{E}[F(k)].\label{eq:stability}
\end{equation}

\subsection{Slicing schemes}\label{ssec:sched}

We can then define four ways of allocating resources to the \gls{vr} users:
\begin{enumerate}
    \item \emph{\gls{if}}: each individual \gls{vr} user is allocated to a different slice that it can fully exploit, and each slice has a constant bandwidth over the next $S$ frames. The bandwidth is then given by:
          \begin{equation}
              B_{\text{IF}}^{(m)}(kS+\ell)=\frac{
              P_m^{-1}(p_s|S,1,kS)+\frac{q_m(t)}{S}}{\eta_mT_{\text{tx}}},
          \end{equation}
          with $\ell\in\{1,\ldots,S\}$. The scheme uses \gls{fdma}, as the bandwidth $B$ is constant over the whole slicing interval. In order to avoid instability, the $q_m(t)$ queued bits need to be considered in the slicing, but they are spread out over the $S$ frames in the slicing period, so as to avoid excessive overprovisioning.
    \item \emph{\gls{io}}: in this case, each user is still assigned to their own individual slice, but there is a finer-grained control over the assignment of bandwidth resources, allowing frame-by-frame control of the bandwidth assignment by using \gls{ofdma}. In this case, the queued bits can be handled in the first frame:
          \begin{equation}
              B_{\text{IO}}^{(m)}(kS+1)=\frac{P_m^{-1}(p_s|1,1,kS) + q_m(t)}{\eta_m T_{\text{tx}}}.
          \end{equation}
          In all subsequent frames, i.e., for $\ell\in\{2,\ldots,S\}$, the bandwidth assignment $B_{\text{IF}}^{(m)}(t)$ is then given by:
          \begin{equation}
              B_{\text{IO}}^{(m)}(kS+\ell)=\frac{P_m^{-1}(p_s|1,\ell,kS)}{\eta_m T_{\text{tx}}}.
          \end{equation}
    \item \emph{\gls{af}}: while the two schemes described above give each user a slice of their own, this scheme performs \gls{fdma}, so it maintains a constant bandwidth throughout, but considers a single slice for the \gls{vr} service. This allows  users with larger than expected frames to exploit the bandwidth left unused by others with smaller than expected frames, but requires another, more fine-grained scheduler to divide the resources among users, which we will describe below. If we consider an oracle prediction, the required bandwidth $B^*(kS)$ to deliver all the generated data is given by:
          \begin{equation}
              B^*(kS)=\sum_{m=1}^M\frac{\frac{q_m(t)}{S}+\sum_{\ell=1}^{S}F^{(m)}(kS+\ell)}{\eta_m T_{\text{tx}}}.
          \end{equation}
          We can split the required bandwidth in two components:
          \begin{equation}
              B^*(kS)=B^*_q(kS)+B^*_f(kS),
          \end{equation}
          where we have:
          \begin{align}
              B^*_q(kS) & =\sum_{m=1}^M\frac{q_m(t)}{S\eta_m T_{\text{tx}}};                                              \\
              B^*_f(kS) & =\sum_{m=1}^M\sum_{\ell=1}^{S}\frac{F^{(m)}(kS+\ell)}{\eta_m T_{\text{tx}}}.\label{eq:B_req_AF}
          \end{align}
          While $B^*_q(kS)$ is a deterministic, known value, as it only depends on the amount of queued bytes for each user, the bandwidth $B^*_f(kS)$ required to transmit future frames is unknown, as the size of these frames is stochastic. The distribution of $B^*_f(kS)$ is given by the convolution of $MS$ Laplace distributions, and the details of its computation are given in the Appendix. If we denote the quantile function of this distribution as $P_{1,\ldots,M}^{-1}(p_s|S,1,t)$, we get, with $\ell\in\{1,\ldots,S\}$:
          \begin{equation}
              B_{\text{AF}}(kS + \ell)=\frac{P_{1,\ldots,M}^{-1}(p_s|S,1,kS)+\sum_{m=1}^M\frac{q_m(t)}{S}}{\eta_m T_{\text{tx}}}.
          \end{equation}
          We assume that users are synchronized, so that frames arrive approximately at the same time: this results in a need-based scheduler delivering the frames from all users approximately at the same time, allocating more bandwidth to users with a larger frame (or a lower spectral efficiency). This is a slight simplification, but we can easily adapt the mechanism for the asynchronous case with limited loss of performance. The choice of a need-based scheduler leaves the decision of setting user rates to flow admission, serving users equitably once they access the system.

    \item \emph{\gls{ao}}: as we did for the individual slicing, we can also create aggregated slices with a finer-grained control of the bandwidth allocation. In the first frame, the queues need to be flushed before new data can be transmitted:
          \begin{equation}
              B_{\text{AF}}(kS + 1)=\frac{P_{1,\ldots,M}^{-1}(p_s|1,1,kS)+\sum_{m=1}^M q_m(t)}{\eta_m T_{\text{tx}}}.
          \end{equation}
          For $\ell\in\{2,\ldots,S\}$, we then have:
          \begin{equation}
              B_{\text{AO}}(kS + \ell)=\frac{P_{1,\ldots,M}^{-1}(p_s|1,\ell,kS)}{\eta_m T_{\text{tx}}}.
          \end{equation}
\end{enumerate}

Naturally, if there is only one user, i.e., $M=1$, the \gls{af} and \gls{ao} slicing schemes are the same as the \gls{if} and \gls{io}, respectively. In the same way, the \gls{fdma} and \gls{ofdma} slicing schemes are equivalent if $S=1$, as the allocation is performed over the shortest possible unit of time, i.e., a single frame period. All slicing schemes ensure the stability of the queue by considering $q_m(t)$ in the bandwidth allocation, inherently ensuring that the condition in~\eqref{eq:stability} is met. The calculation of the aggregated traffic distribution is given in the Appendix.

\begin{figure*}[t!]
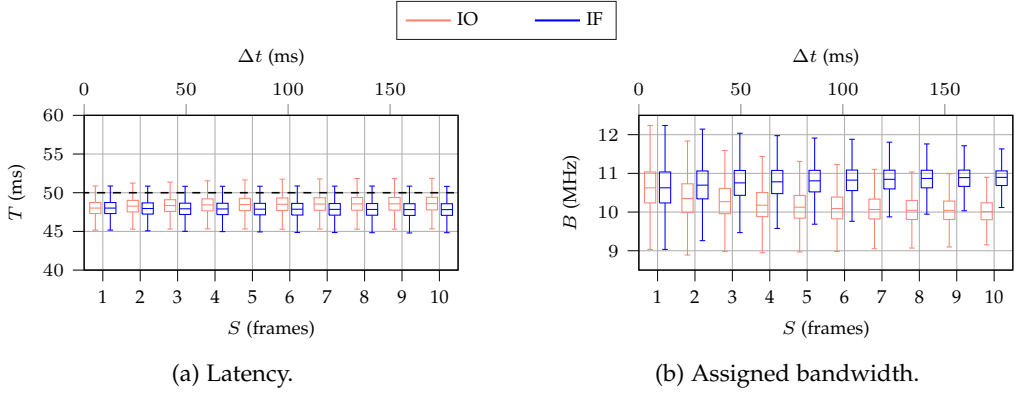

    \centering
    \begin{subfigure}[b]{\linewidth}
        \centering
\begin{tikzpicture}

\begin{axis}[
    width=0,
    height=0,
    at={(0,0)},
    scale only axis,
    xmin=0,
    xmax=0,
    xtick={},
    ymin=0,
    ymax=0,
    ytick={},
    axis background/.style={fill=white},
    legend style={legend cell align=center, align=center, draw=white!15!black, font=\scriptsize, at={(0, 0)}, anchor=center, /tikz/every even column/.append style={column sep=2em}},
    legend columns=10,
]
\addplot [thick, color2]
table {%
0 1
};
\addlegendentry{\gls{io}}
\addplot [thick, color6]
table {%
0 1
};
\addlegendentry{\gls{if}}

\end{axis}

\end{tikzpicture}
    \end{subfigure}
    \\
    \begin{subfigure}[b]{.4\linewidth}
        \centering
        \input{./img/single_Quantile_latency_granularity.tex}
        \caption{Latency.}
        \label{fig:quant_latbox}
    \end{subfigure}
    \begin{subfigure}[b]{.4\linewidth}
        \centering
        \input{./img/single_Quantile_rate_granularity.tex}
        \caption{Assigned bandwidth.}
        \label{fig:quant_schbox}
    \end{subfigure}
    \caption{Boxplot of slicing performance for \gls{if} and \gls{io} for a single user.}
    \label{fig:latsch_gran}
\end{figure*}

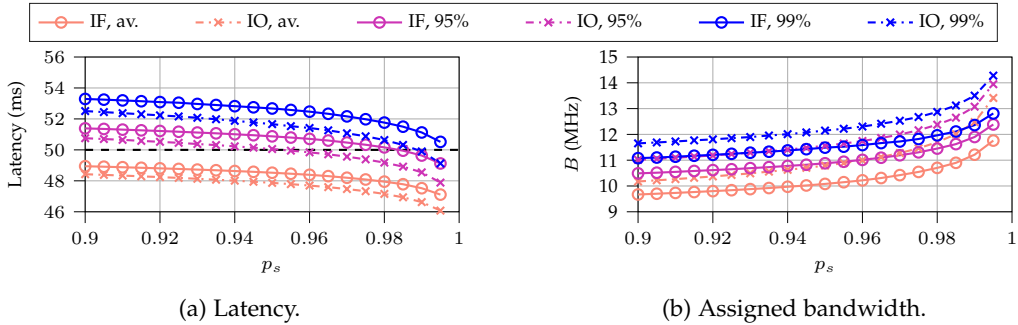
\begin{figure*}[t!]
    \centering
    \begin{subfigure}[b]{\linewidth}
        \centering
\begin{tikzpicture}

\begin{axis}[
    width=0,
    height=0,
    at={(0,0)},
    scale only axis,
    xmin=0,
    xmax=0,
    xtick={},
    ymin=0,
    ymax=0,
    ytick={},
    axis background/.style={fill=white},
    legend style={legend cell align=center, align=center, draw=white!15!black, font=\scriptsize, at={(0, 0)}, anchor=center, /tikz/every even column/.append style={column sep=2em}},
    legend columns=10,
]
\addplot [thick, color2, mark=o, mark options={solid}]
table {%
0 1
};
\addlegendentry{\gls{if}, av.}
\addplot [thick, dashdotted, color2, mark=x, mark options={solid}]
table {%
0 1
};
\addlegendentry{\gls{io}, av.}

\addplot [thick, color4, mark=o, mark options={solid}]
table {%
0 1
};
\addlegendentry{\gls{if}, 95\%}
\addplot [thick, dashdotted, color4, mark=x, mark options={solid}]
table {%
0 1
};
\addlegendentry{\gls{io}, 95\%}

\addplot [thick, color6, mark=o, mark options={solid}]
table {%
0 1
};
\addlegendentry{\gls{if}, 99\%}
\addplot [thick, dashdotted, color6, mark=x, mark options={solid}]
table {%
0 1
};
\addlegendentry{\gls{io}, 99\%}

\end{axis}

\end{tikzpicture}
    \end{subfigure}
    \\
    \begin{subfigure}[b]{.4\linewidth}
        \centering
\begin{tikzpicture}

\begin{axis}[
width=\sfwidth,
height=\sfheight,
legend cell align={left},
legend style={at={(0.995,0.99)}, anchor=north east, fill opacity=0.8, legend columns=2, draw opacity=1, text opacity=1, font=\tiny, draw=white!80!black},
tick align=outside,
tick pos=left,
x grid style={white!69.0196078431373!black},
xlabel={$p_s$},
xmin=0.9, xmax=1,
xtick style={color=black},
y grid style={white!69.0196078431373!black},
ylabel={Latency (ms)},
ymin=46, ymax=56,
ytick style={color=black},
xmajorgrids,
ymajorgrids
]
\addplot[thick, black, dashed, forget plot]
table{
0.9 50
1 50
};

\addplot [thick, color2, mark=o, mark options={solid}]
table {%
0.9 48.9434451878751
0.905 48.9119335481849
0.91 48.8785658321278
0.915 48.8427382101195
0.92 48.8060032561495
0.925 48.7682986378749
0.93 48.7258087082252
0.935 48.6827406069212
0.94 48.6325166754796
0.945 48.5823860347284
0.95 48.5235993273326
0.955 48.4583036495305
0.96 48.3889834606261
0.965 48.3032439741635
0.97 48.1985773279137
0.975 48.0858531406434
0.98 47.9474151944356
0.985 47.7794130039347
0.99 47.5224540384851
0.995 47.1096120973911
};

\addplot [thick, dashdotted, color2, mark=x, mark options={solid}]
table {%
0.9 48.4274058089461
0.905 48.383995559957
0.91 48.3394483043331
0.915 48.2908346804697
0.92 48.2415214012092
0.925 48.1883933785613
0.93 48.1320180389285
0.935 48.071920949081
0.94 48.0095170651241
0.945 47.941057090573
0.95 47.8670054085434
0.955 47.7855453764788
0.96 47.6947127617821
0.965 47.590993530645
0.97 47.4692636261129
0.975 47.3275200245883
0.98 47.1536647400488
0.985 46.9352245801617
0.99 46.62545204108
0.995 46.0751318140286
};

\addplot [thick, color4, mark=o, mark options={solid}]
table {%
0.9 51.3890200755532
0.905 51.3516482656556
0.91 51.3079514278029
0.915 51.2627372873996
0.92 51.2200369166681
0.925 51.1717134923875
0.93 51.1205376713639
0.935 51.0690150732529
0.94 51.0088141381358
0.945 50.9461090493165
0.95 50.8740653515794
0.955 50.7958879148251
0.96 50.7092595027843
0.965 50.6028258449791
0.97 50.4715473488762
0.975 50.3307983306798
0.98 50.1609155678525
0.985 49.9481887608963
0.99 49.6344423298315
0.995 49.1218274581287
};

\addplot [thick, dashdotted, color4, mark=x, mark options={solid}]
table {%
0.9 50.7531780941165
0.905 50.7004441024325
0.91 50.6467843621279
0.915 50.5803303707741
0.92 50.5191999067355
0.925 50.4552309823376
0.93 50.3780173842393
0.935 50.3099871794875
0.94 50.2286066144875
0.945 50.1433832809413
0.95 50.0531547257892
0.955 49.9499275418867
0.96 49.8397175768552
0.965 49.7123638279011
0.97 49.5610943788527
0.975 49.3862232790596
0.98 49.1770039521723
0.985 48.9160053227663
0.99 48.5479709822747
0.995 47.8849691828008
};

\addplot [thick, color6, mark=o, mark options={solid}]
table {%
0.9 53.289228063272
0.905 53.246054983565
0.91 53.1959826280621
0.915 53.1436328296007
0.92 53.0879262797217
0.925 53.0468205541121
0.93 52.9667477970688
0.935 52.8985710802511
0.94 52.8233565213234
0.945 52.7499756400511
0.95 52.6743742407101
0.955 52.5790085727179
0.96 52.4690320188458
0.965 52.3311893487422
0.97 52.1639401465546
0.975 51.9780868077661
0.98 51.7625097344299
0.985 51.4977191975167
0.99 51.1248644393249
0.995 50.5164876771948
};

\addplot [thick, dashdotted, color6, mark=x, mark options={solid}]
table {%
0.9 52.5066279588505
0.905 52.4413446417971
0.91 52.3717844465064
0.915 52.3014216233697
0.92 52.2303263241509
0.925 52.159355087218
0.93 52.0734324316508
0.935 51.9798637068728
0.94 51.8761403304192
0.945 51.7714760016485
0.95 51.6641963931443
0.955 51.5314304707428
0.96 51.4130615202165
0.965 51.2620178293881
0.97 51.0714930242855
0.975 50.8768052518118
0.98 50.6343292791249
0.985 50.3177679289324
0.99 49.8990454474175
0.995 49.1251910502197
};

\end{axis}

\end{tikzpicture}
        \caption{Latency.}
        \label{fig:perc_lat_ps}
    \end{subfigure}
    \begin{subfigure}[b]{.4\linewidth}
        \centering
\begin{tikzpicture}

\begin{axis}[
width=\sfwidth,
height=\sfheight,
legend cell align={left},
legend style={at={(0.01,0.98)}, anchor=north west, fill opacity=0.8, legend columns=2, draw opacity=1, text opacity=1, font=\tiny, draw=white!80!black},
tick align=outside,
tick pos=left,
x grid style={white!69.0196078431373!black},
xlabel={$p_s$},
xmin=0.9, xmax=1,
xtick style={color=black},
y grid style={white!69.0196078431373!black},
ylabel={$B$ (MHz)},
ymin=45, ymax=75,
ytick={45,50,55,60,65,70,75},
yticklabels={9,10,11,12,13,14,15},
ytick style={color=black},
xmajorgrids,
ymajorgrids
]
\addplot [thick, color2, mark=o, mark options={solid}]
table {%
0.9 48.3515626879969
0.905 48.4990966641216
0.91 48.6566917403121
0.915 48.8271831726059
0.92 49.0029659040077
0.925 49.1841883229669
0.93 49.3907933505646
0.935 49.6017602437092
0.94 49.8504491510335
0.945 50.1011471967127
0.95 50.3990342601646
0.955 50.7329896993246
0.96 51.0925398835215
0.965 51.5435692687682
0.97 52.1057741956214
0.975 52.7251857904483
0.98 53.5090001328513
0.985 54.489620487366
0.99 56.0654401419261
0.995 58.7989868071364
};

\addplot [thick, dashdotted, color2, mark=x, mark options={solid}]
table {%
0.9 50.8728835137964
0.905 51.0976717431031
0.91 51.3309523871338
0.915 51.5879271630387
0.92 51.851200523561
0.925 52.1385224829881
0.93 52.4466427152764
0.935 52.779959351651
0.94 53.1306381145937
0.945 53.5209165462969
0.95 53.9500408963801
0.955 54.4300466046065
0.96 54.97709284743
0.965 55.6157442567093
0.97 56.3850380468565
0.975 57.3095571739889
0.98 58.4869547097448
0.985 60.0362091872699
0.99 62.385110282659
0.995 67.0599402684868
};

\addplot [thick, color4, mark=o, mark options={solid}]
table {%
0.9 52.449114381913
0.905 52.5744378512364
0.91 52.7347670357867
0.915 52.9087742949143
0.92 53.0756852599174
0.925 53.2407909985628
0.93 53.4566157333879
0.935 53.6504151515545
0.94 53.8886844113934
0.945 54.1171315702239
0.95 54.3973911260326
0.955 54.6994348416898
0.96 55.0565187125861
0.965 55.4782002285368
0.97 55.9549419761067
0.975 56.5235199879135
0.98 57.2532761530286
0.985 58.1174891180878
0.99 59.4959818527302
0.995 61.9226635331717
};

\addplot [thick, dashdotted, color4, mark=x, mark options={solid}]
table {%
0.9 55.0673391914723
0.905 55.2637351416627
0.91 55.4723263581524
0.915 55.697396913107
0.92 55.919504226601
0.925 56.1881139376373
0.93 56.4585100850783
0.935 56.7551474759855
0.94 57.0594422683339
0.945 57.4139191928001
0.95 57.8006145767596
0.955 58.226460038326
0.96 58.7165775232064
0.965 59.3008733333385
0.97 59.9757980335217
0.975 60.8055295657037
0.98 61.8824962371782
0.985 63.2452020866162
0.99 65.3773541395129
0.995 69.7285656755256
};

\addplot [thick, color6, mark=o, mark options={solid}]
table {%
0.9 55.397833237834
0.905 55.5141180201058
0.91 55.6745901640144
0.915 55.8612908948967
0.92 56.0328195353818
0.925 56.2357987883009
0.93 56.4515245536688
0.935 56.6430317278639
0.94 56.8902224342415
0.945 57.1365874664925
0.95 57.4348870710356
0.955 57.6200910861634
0.96 57.9330700086568
0.965 58.3173545413849
0.97 58.6264724388547
0.975 59.107539572881
0.98 59.8170074732172
0.985 60.5566176731416
0.99 61.8114808029326
0.995 64.0618257680548
};
\addplot [thick, dashdotted, color6, mark=x, mark options={solid}]
table {%
0.9 58.2634256206565
0.905 58.4373098027008
0.91 58.6325955869072
0.915 58.8435232380168
0.92 59.0402545503159
0.925 59.2705637130443
0.93 59.4876595809513
0.935 59.7851886908283
0.94 60.0421418465836
0.945 60.3865646859817
0.95 60.7427790168431
0.955 61.0959032526491
0.96 61.5201688702556
0.965 62.0658921127619
0.97 62.6526389007377
0.975 63.3674231312404
0.98 64.3482057533011
0.985 65.6043572299744
0.99 67.5153866097659
0.995 71.3940408782368
};

\end{axis}

\end{tikzpicture}
        \caption{Assigned bandwidth.}
        \label{fig:perc_sch_ps}
    \end{subfigure}
    \caption{Average and worst-case percentiles of the latency and assigned bandwidth of a single user as a function of the quantile $p_s$.}
    \label{fig:ps_perc}
\end{figure*}

\section{Simulation Results}\label{sec:results}

In the following, we run a simulation on a system implementing the slicing schemes we described in the previous section, analyzing the two fundamental \glspl{kpi} of the system: the \gls{mtp} latency and the bandwidth $B$ reserved to the \gls{vr} users. Ideally, a system should be able to maintain the \gls{mtp} latency below the required threshold while limiting the required bandwidth. The main parameters of the scenario are listed in Table~\ref{tab:latency}, and will be used in all simulations, unless stated otherwise. We chose $T_{\max}=50$~ms, which is consistent with the relevant literature, though looser than in the IEEE standard: as the application we used for the measuring has $\Delta_u=7$~ms (on average) and $\varphi=60$~\gls{fps}, even an instantaneous transmission would incur an \gls{mtp} latency over 20~ms in the worst case. The stricter deadline set by the IEEE standard is then impossible to reach with the considered application, and we chose a looser but still realistic deadline, leaving the fulfillment of the more demanding one to future work with more powerful \gls{xr} applications.

We also considered $\tau_p=5$~ms and $\tau_r=5$~ms, considering a powerful Cloud \gls{vr} server located relatively close to the user. The final parameters are close to the 3GPP recommendation~\cite{3gpp.38.838}, which recommends a rate $R=30$~Mb/s and $\varphi=60$~\gls{fps}, although our \gls{dl} latency budget is slightly looser, as the \gls{bs} has 11.3~ms to stream each frame, while 3GPP specifies 10~ms as the target.

\begin{table}[b]
    \centering
    \scriptsize
    \caption{Basic scenario parameters.}
    \label{tab:latency}
    \begin{tabular}{c|c}
        \toprule
        Parameter     & Value        \\
        \midrule
        $T_{\max}$    & 50~ms        \\
        $\Delta_u$    & 7~ms         \\
        $\tau_p$      & 5~ms         \\
        $\tau_r$      & 5~ms         \\
        $p_s$         & 0.95         \\
        $S$           & 6~frames     \\
        $N$           & 6~frames     \\
        $\tau$        & 1~frame      \\
        $R$           & 30~Mb/s      \\
        $\varphi$     & 60~\gls{fps} \\
        Video content & Virus Popper \\
        \bottomrule
    \end{tabular}
\end{table}

\subsection{Single User}

We can now examine the simulation results for a single user. As we remarked in the previous section, the \gls{io} scheme can reduce the jitter by having a more fine-grained prediction, as each frame will be allocated enough resources to be transmitted with probability $p_s$.
On the other hand, the \gls{if} scheme has a rougher prediction, with consequently higher jitter, but will waste fewer network resources, as it can allow larger frames to be compensated by smaller ones before and after them. Both models are realistic, as they work under different assumptions: in the first case, the resources that are allocated for each frame need to be over both time and frequency, while the second case gives the slice a constant bandwidth over the slicing interval, which is the most common slicing model in the literature.

We can then look at the slicing schemes' performance as a function of $S$, setting $p_s=0.95$ and $N=6$: Fig.~\ref{fig:latsch_gran} shows boxplots of the latency and assigned bandwidth for \gls{if} and \gls{io}. Fig.~\ref{fig:quant_latbox} clearly shows that, while the slicing granularity has a limited effect on \gls{io}, the
lower precision of \gls{if} means that the longer the slicing interval, the higher the average latency, and
the worst-case latency, represented by the upper whisker of the boxplots, increases even more.
On the other hand, as
Fig.~\ref{fig:quant_schbox} shows, the bandwidth required by \gls{if} decreases as $S$ grows, while
the average bandwidth required by the \gls{io} algorithm remains roughly constant irrespective of the value of $S$, but always higher than the bandwidth used by \gls{if}.

This behavior is to be expected, as the errors in frame prediction can compensate over a longer window, but comes at the cost of a higher latency. Naturally, the choice between the two models depends not only on the desired point in the trade-off between \gls{qos} and resource efficiency, but also on the capabilities of the underlying system: state-of-the-art slicing frameworks often consider a period $\Delta t=100$~ms, which would correspond to $S=6$ frames, and the granularity of the slicing over time and frequency will dictate whether \gls{io} is even an option.

\begin{figure*}[t!]
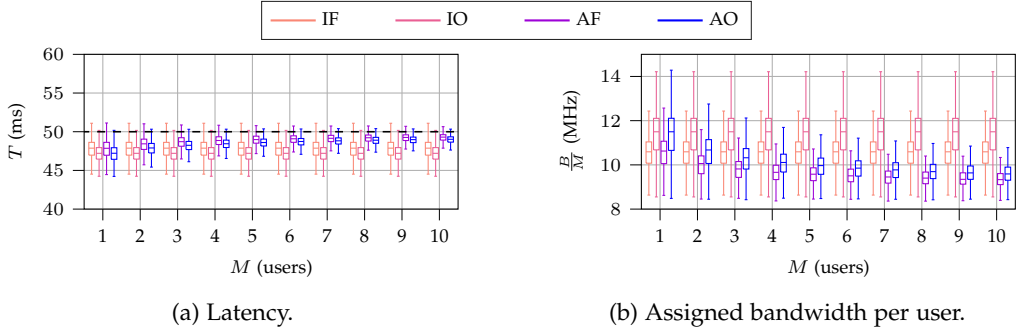

    \centering
    \begin{subfigure}[b]{\linewidth}
        \centering
\begin{tikzpicture}

\begin{axis}[
    width=0,
    height=0,
    at={(0,0)},
    scale only axis,
    xmin=0,
    xmax=0,
    xtick={},
    ymin=0,
    ymax=0,
    ytick={},
    axis background/.style={fill=white},
    legend style={legend cell align=center, align=center, draw=white!15!black, font=\scriptsize, at={(0, 0)}, anchor=center, /tikz/every even column/.append style={column sep=2em}},
    legend columns=10,
]
\addplot [thick, color2]
table {%
0 1
};
\addlegendentry{\gls{if}}
\addplot [thick, color3]
table {%
0 1
};
\addlegendentry{\gls{io}}
\addplot [thick, color5]
table {%
0 1
};
\addlegendentry{\gls{af}}
\addplot [thick, color6]
table {%
0 1
};
\addlegendentry{\gls{ao}}

\end{axis}

\end{tikzpicture}
    \end{subfigure}
    \\
    \begin{subfigure}[b]{.4\linewidth}
        \centering
        \input{./img/multi_latency_users.tex}
        \caption{Latency.}
        \label{fig:lat_multi}
    \end{subfigure}
    \begin{subfigure}[b]{.4\linewidth}
        \centering
        \input{./img/multi_rate_users.tex}
        \caption{Assigned bandwidth per user.}
        \label{fig:sch_multi}
    \end{subfigure}
    \caption{Boxplot of the latency and per-user bandwidth as a function of the number of users.}
    \label{fig:multiuser}
\end{figure*}

\begin{figure*}[t!]
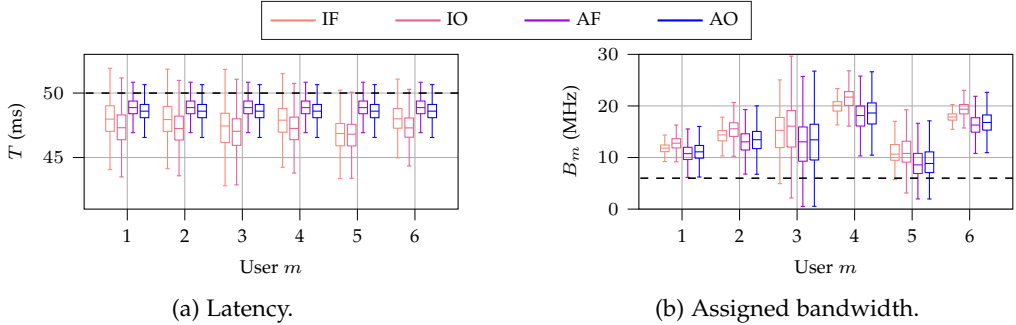

    \centering
    \begin{subfigure}[b]{\linewidth}
        \centering
\begin{tikzpicture}

\begin{axis}[
    width=0,
    height=0,
    at={(0,0)},
    scale only axis,
    xmin=0,
    xmax=0,
    xtick={},
    ymin=0,
    ymax=0,
    ytick={},
    axis background/.style={fill=white},
    legend style={legend cell align=center, align=center, draw=white!15!black, font=\scriptsize, at={(0, 0)}, anchor=center, /tikz/every even column/.append style={column sep=2em}},
    legend columns=10,
]
\addplot [thick, color2]
table {%
0 1
};
\addlegendentry{\gls{if}}
\addplot [thick, color3]
table {%
0 1
};
\addlegendentry{\gls{io}}
\addplot [thick, color5]
table {%
0 1
};
\addlegendentry{\gls{af}}
\addplot [thick, color6]
table {%
0 1
};
\addlegendentry{\gls{ao}}

\end{axis}

\end{tikzpicture}
    \end{subfigure}
    \\
    \begin{subfigure}[b]{.4\linewidth}
        \centering
        \input{./img/box_latency_per_user.tex}
        \caption{Latency.}
        \label{fig:lat_peruser}
    \end{subfigure}
    \begin{subfigure}[b]{.4\linewidth}
        \centering
        \input{./img/box_rate_per_user.tex}
        \caption{Assigned bandwidth.}
        \label{fig:sch_peruser}
    \end{subfigure}
    \caption{Boxplot of the latency and bandwidth for each user.}
    \label{fig:per_user}
\end{figure*}

It is also possible to simply increase the value of $C(t)$, e.g., by increasing $p_s$, in the \gls{if} scheme to match the \gls{io} performance in terms of latency, but \gls{if} will always be less efficient for the same latency target. Fig.~\ref{fig:ps_perc} shows the slicing performance as a function of the value of $p_s$. Naturally, a higher $p_s$ means a more conservative prediction of getting larger frames, which reduces the latency but increases the bandwidth requirements.
The closer we get to 1, the more increasing $p_s$ affects the latency, with a correspondingly larger increase in the bandwidth that is reserved to the \gls{vr} flow.
We can also notice that \gls{if} requires a much higher value of $p_s$ to get the same performance as \gls{io} in terms of latency.
A sensible example is to target a latency of one inter-frame interval, i.e., $\varphi^{-1}=16.67$~ms (the dashed line in \cref{fig:perc_lat_ps}), with a probability of 0.95 (the pink lines in \cref{fig:ps_perc}).
We notice that to meet this requirement, a value of $p_s\geq0.96$ has to be chosen for the \gls{io} scheme, but the same requirement can only be fulfilled if $p_s\geq0.99$ using \gls{if}.
This corresponds to an average assigned bandwidth of at least 7.5~MHz for \gls{io}, but 7.64~MHz for \gls{if}. While the difference is not very significant, and the \gls{if} scheme can be used without a big performance loss, choosing the correct value of $p_s$ to compensate for the slicing scheme's optimism is not simple, particularly in more complex network scenarios, while it is relatively straightforward for \gls{io}.

\subsection{Multiple Users}

We can now consider a more complex scenario, in which multiple \gls{vr} users are served by the same \gls{bs}. The first thing we need to understand is the impact of an increased number of users on the latency and allocation of resources. In this case, we will set up a scenario in which all users have the same spectral efficiency $\eta=5$~b/s/Hz, and stream the same content (i.e., the \emph{Virus popper} trace at 60~\gls{fps} and 30~Mb/s), although starting with a random offset. This is to ensure that the scenario is as uniform as possible, and the only variable is the number $M$ of users in the system. Furthermore, as we stated in Sec.~\ref{ssec:sched}, we assume that all \gls{vr} users are synchronized, i.e., frames from all users come at the same time: this is a slight simplification, which does not have a significant effect on performance.

Fig.~\ref{fig:multiuser} shows the latency and assigned bandwidth for the scenario: if we look at the latency boxplot in Fig.~\ref{fig:lat_multi}, we can note that the average latency for the two aggregated schemes increases with $M$, but the higher percentiles actually decrease as $M$ increases: this is because, as aggregating more users leads to errors in the frame size prediction compensating each other, the resource allocation can be more precise, aiming at satisfying the $\varphi^{-1}$ latency requirement, but not trying to go far below that. As the number of users grows, the distribution of the latency will tend towards deterministically achieving $\varphi^{-1}$ with limited jitter. This is a significant advantage with respect to individual slicing, which is compounded by the lower resource use, as Fig.~\ref{fig:sch_multi} shows: as the number of users increases, the bandwidth allocated to each user remains the same for individual slicing schemes, while the aggregate schemes can reduce the total amount of resources. As for the latency, this is because independent variations in the frame size compensate each other, leading to a lower overall uncertainty on the total frame size and a better resource provisioning. Finally, as for the single user case, using \gls{ofdma} can slightly reduce the latency, overshooting the deadline less often, but at the cost of a higher resource utilization: the same error compensation effect that we mentioned across different users can be exploited over different frames, at the cost of a higher latency violation probability.

\begin{table}[b]
    \centering
    \scriptsize
    \caption{Multi-user scenario parameters.}
    \label{tab:scenario}
    \begin{tabular}{c|ccccc}
        \toprule
        User $m$ & VR content   & $R_m$ (Mb/s) & $\eta_m$ (b/s/Hz) & $\bar{B}_m (MHz)$ \\
        \midrule
        1        & Virus Popper & 10           & 1.5               & 6.67              \\
        2        & Cities       & 20           & 2.5               & 8                 \\
        3        & Minecraft    & 30           & 3                 & 10                \\
        4        & Tour         & 40           & 3.5               & 11.43             \\
        5        & Minecraft    & 50           & 5.5               & 9.09              \\
        6        & Virus Popper & 40           & 4                 & 10                \\
        \bottomrule
    \end{tabular}
\end{table}

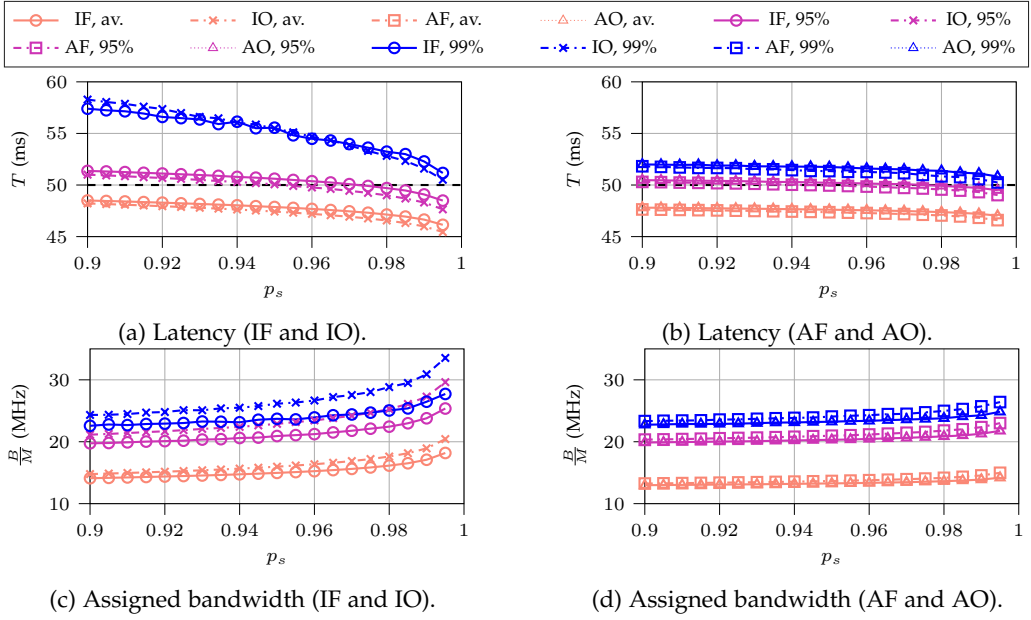
\begin{figure*}[t!]
    \centering
    \begin{subfigure}[b]{\linewidth}
        \centering
\begin{tikzpicture}

\begin{axis}[
    width=0,
    height=0,
    at={(0,0)},
    scale only axis,
    xmin=0,
    xmax=0,
    xtick={},
    ymin=0,
    ymax=0,
    ytick={},
    axis background/.style={fill=white},
    legend style={legend cell align=center, align=center, draw=white!15!black, font=\scriptsize, at={(0, 0)}, anchor=center, legend columns=6, /tikz/every even column/.append style={column sep=2em}}  
]
\addplot [thick, color2, mark=o, mark options={solid}]
table {%
0 1
};
\addlegendentry{\gls{if}, av.}

\addplot [thick, dashdotted, color2, mark=x, mark options={solid}]
table {%
0 1
};
\addlegendentry{\gls{io}, av.}

\addplot [thick, dashdotted, color2, mark=square, mark options={solid}]
table {%
0 1
};
\addlegendentry{\gls{af}, av.}

\addplot [densely dotted, color2, mark=triangle, mark options={solid}]
table {%
0 1
};
\addlegendentry{\gls{ao}, av.}

\addplot [thick, color4, mark=o, mark options={solid}]
table {%
0 1
};
\addlegendentry{\gls{if}, 95\%}

\addplot [thick, dashdotted, color4, mark=x, mark options={solid}]
table {%
0 1
};
\addlegendentry{\gls{io}, 95\%}

\addplot [thick, dashdotted, color4, mark=square, mark options={solid}]
table {%
0 1
};
\addlegendentry{\gls{af}, 95\%}

\addplot [densely dotted, color4, mark=triangle, mark options={solid}]
table {%
0 1
};
\addlegendentry{\gls{ao}, 95\%}

\addplot [thick, color6, mark=o, mark options={solid}]
table {%
0 1
};
\addlegendentry{\gls{if}, 99\%}

\addplot [thick, dashdotted, color6, mark=x, mark options={solid}]
table {%
0 1
};
\addlegendentry{\gls{io}, 99\%}

\addplot [thick, dashdotted, color6, mark=square, mark options={solid}]
table {%
0 1
};
\addlegendentry{\gls{af}, 99\%}

\addplot [densely dotted, color6, mark=triangle, mark options={solid}]
table {%
0 1
};
\addlegendentry{\gls{ao}, 99\%}

\end{axis}

\end{tikzpicture}
    \end{subfigure}
    \\
    \begin{subfigure}[b]{.4\linewidth}
        \centering
\begin{tikzpicture}

\begin{axis}[
width=\sfwidth,
height=\sfheight,
legend cell align={left},
legend style={at={(0.995,0.99)}, anchor=north east, fill opacity=0.8, legend columns=2, draw opacity=1, text opacity=1, font=\tiny, draw=white!80!black},
tick align=outside,
tick pos=left,
x grid style={white!69.0196078431373!black},
xlabel={$p_s$},
xmin=0.9, xmax=1,
xtick style={color=black},
y grid style={white!69.0196078431373!black},
ylabel={$T$ (ms)},
ymin=45, ymax=60,
ytick style={color=black},
xmajorgrids,
ymajorgrids
]
\addplot[thick, black, dashed, forget plot]
table{
0.9 50
1 50
};

\addplot [thick, color2, mark=o, mark options={solid}]
table {%
0.9 48.4993967606938
0.905 48.4512572865712
0.91 48.4066011589986
0.915 48.3452178470591
0.92 48.2924070644586
0.925 48.2275622685611
0.93 48.1657337054581
0.935 48.1014407403556
0.94 48.0362984604818
0.945 47.9495343593321
0.95 47.8706407869446
0.955 47.7830572828011
0.96 47.6843074810456
0.965 47.5762152219879
0.97 47.4493954131068
0.975 47.3147405560162
0.98 47.1376357768856
0.985 46.9407851696833
0.99 46.6474472073545
0.995 46.1422670267385
};

\addplot [thick, dashdotted, color2, mark=x, mark options={solid}]
table {%
0.9 48.219225120135
0.905 48.1514631909894
0.91 48.0914568817451
0.915 48.0189957393767
0.92 47.9468186869648
0.925 47.873182832024
0.93 47.7941463143859
0.935 47.7113878753223
0.94 47.6220890519267
0.945 47.5300259161482
0.95 47.4281890353453
0.955 47.3175857502394
0.96 47.2025138047287
0.965 47.0710144426021
0.97 46.9263691680114
0.975 46.7603494336191
0.98 46.5618930974224
0.985 46.317704113384
0.99 45.9909631911188
0.995 45.4421967672812
};

\addplot [thick, color4, mark=o, mark options={solid}]
table {%
0.9 51.3663087621467
0.905 51.314816550788
0.91 51.2527713276531
0.915 51.1842421219488
0.92 51.1157990645052
0.925 51.0425595028718
0.93 50.9721308687491
0.935 50.8881332472399
0.94 50.7872917680119
0.945 50.7103799968431
0.95 50.6095033597256
0.955 50.4904436380072
0.96 50.3712861612944
0.965 50.2463191330911
0.97 50.0851517417894
0.975 49.9074624550499
0.98 49.7024253752332
0.985 49.4697017646064
0.99 49.0880910649538
0.995 48.490188464115
};

\addplot [thick, dashdotted, color4, mark=x, mark options={solid}]
table {%
0.9 51.0031644495656
0.905 50.9333458391569
0.91 50.8522793698022
0.915 50.7583408184099
0.92 50.6810683860086
0.925 50.5941632320824
0.93 50.4964060606865
0.935 50.3883955056079
0.94 50.2996021018163
0.945 50.1855738856642
0.95 50.067890961209
0.955 49.9085953325095
0.96 49.7625810519875
0.965 49.6204070052878
0.97 49.4374534571105
0.975 49.2383558109252
0.98 49.0025812035601
0.985 48.6933067140078
0.99 48.3070118225749
0.995 47.6395448080156
};

\addplot [thick, color6, mark=o, mark options={solid}]
table {%
0.9 57.396877412022
0.905 57.2444169565566
0.91 57.1456565711616
0.915 56.9281443862243
0.92 56.6001158804894
0.925 56.482101119172
0.93 56.3563116718018
0.935 55.9076379935472
0.94 56.1346649777122
0.945 55.5059413610838
0.95 55.5568128602914
0.955 54.819293749354
0.96 54.4845780149685
0.965 54.3127088394299
0.97 53.969075964889
0.975 53.6165829768005
0.98 53.2351438030683
0.985 52.9947533252413
0.99 52.2839937499117
0.995 51.1669098989328
};

\addplot [thick, dashdotted, color6, mark=x, mark options={solid}]
table {%
0.9 58.2833293056942
0.905 58.0469683644163
0.91 57.8699854329472
0.915 57.5975499393701
0.92 57.3658709996721
0.925 56.9795581267628
0.93 56.6005504164848
0.935 56.4698982296579
0.94 56.0339801093436
0.945 55.8627017921521
0.95 55.4805730701777
0.955 55.1102586405702
0.96 54.674669848024
0.965 54.3438194237356
0.97 53.8286595230928
0.975 53.3094822725401
0.98 52.8249830545991
0.985 52.3356710002016
0.99 51.5879571246481
0.995 50.511512686903
};

\end{axis}

\end{tikzpicture}
        \caption{Latency (IF and IO).}
        \label{fig:perc_lat_ind}
    \end{subfigure}
    \begin{subfigure}[b]{.4\linewidth}
        \centering
\begin{tikzpicture}

\begin{axis}[
width=\sfwidth,
height=\sfheight,
legend cell align={left},
legend style={at={(0.995,0.99)}, anchor=north east, fill opacity=0.8, legend columns=2, draw opacity=1, text opacity=1, font=\tiny, draw=white!80!black},
tick align=outside,
tick pos=left,
x grid style={white!69.0196078431373!black},
xlabel={$p_s$},
xmin=0.9, xmax=1,
xtick style={color=black},
y grid style={white!69.0196078431373!black},
ylabel={$T$ (ms)},
ymin=45, ymax=60,
ytick style={color=black},
xmajorgrids,
ymajorgrids
]
\addplot[thick, black, dashed, forget plot]
table{
0.9 50
1 50
};

\addplot [thick, color2, mark=triangle, mark options={solid}]
table {%
0.9 47.8082865902645
0.905 47.793374984777
0.91 47.7763418662186
0.915 47.7604449415828
0.92 47.7421917039426
0.925 47.7233585865141
0.93 47.7036088861781
0.935 47.6820372087837
0.94 47.6599444536111
0.945 47.6349582615418
0.95 47.60836351442
0.955 47.5832154918819
0.96 47.5507755684835
0.965 47.5154045498715
0.97 47.4801120121263
0.975 47.4286776829461
0.98 47.3759702197356
0.985 47.3024058828823
0.99 47.2087033928637
0.995 47.0280313415421
};

\addplot [thick, dashdotted, color2, mark=square, mark options={solid}]
table {%
0.9 47.6329708148967
0.905 47.611924168862
0.91 47.5890735688037
0.915 47.5658084542628
0.92 47.5414653740709
0.925 47.5153757667726
0.93 47.4893350363014
0.935 47.4601964850903
0.94 47.43011752871
0.945 47.397483527854
0.95 47.3633276504349
0.955 47.3249178421564
0.96 47.2838853421046
0.965 47.2362741367742
0.97 47.1840074713765
0.975 47.1221127654267
0.98 47.0497293607396
0.985 46.9570230129583
0.99 46.8284112307587
0.995 46.5972534036987
};

\addplot [thick, color4, mark=triangle, mark options={solid}]
table {%
0.9 50.4721601779016
0.905 50.4554424515288
0.91 50.4275642355232
0.915 50.4179454370536
0.92 50.3980946341679
0.925 50.3794268949335
0.93 50.3533119050851
0.935 50.2961667434017
0.94 50.2962552013622
0.945 50.2549882378983
0.95 50.212072069331
0.955 50.1925010557893
0.96 50.1460715210166
0.965 50.1022566480183
0.97 50.0899627094049
0.975 49.9991312809638
0.98 49.9209511607152
0.985 49.8184526686448
0.99 49.7225970097165
0.995 49.5008677972292
};

\addplot [thick, dashdotted, color4, mark=square, mark options={solid}]
table {%
0.9 50.3018377520313
0.905 50.2767689999295
0.91 50.2380766243729
0.915 50.2129465865287
0.92 50.1827530383513
0.925 50.1587989317823
0.93 50.1248045337414
0.935 50.0738844777215
0.94 50.0318214515169
0.945 49.9981565358115
0.95 49.955090094116
0.955 49.9068612021716
0.96 49.8510726832466
0.965 49.788254737543
0.97 49.7237936191262
0.975 49.66486766058
0.98 49.5649004308609
0.985 49.4619223039053
0.99 49.3154308371139
0.995 49.026452135704
};

\addplot [thick, color6, mark=triangle, mark options={solid}]
table {%
0.9 51.9956611351468
0.905 51.9762585384214
0.91 51.9504118191555
0.915 51.9648938085015
0.92 51.8844448444767
0.925 51.8894578045609
0.93 51.8162444545107
0.935 51.8029262025239
0.94 51.785588221722
0.945 51.7755226429667
0.95 51.7491933565912
0.955 51.678858561813
0.96 51.6015284542826
0.965 51.5527874955483
0.97 51.5169472927976
0.975 51.4111949427337
0.98 51.3760045369213
0.985 51.2026537357214
0.99 51.1171390933954
0.995 50.8282387046447
};

\addplot [thick, dashdotted, color6, mark=square, mark options={solid}]
table {%
0.9 51.8185150602445
0.905 51.7596372035503
0.91 51.7064009075724
0.915 51.6589991616027
0.92 51.6553055731549
0.925 51.629546955921
0.93 51.5324912134664
0.935 51.5193926502744
0.94 51.445020103783
0.945 51.3649547724656
0.95 51.3513841740916
0.955 51.2780187090643
0.96 51.2668943369325
0.965 51.2153032819129
0.97 51.1468578789953
0.975 50.9786789664335
0.98 50.8366356525939
0.985 50.764212700154
0.99 50.5741558079627
0.995 50.2474687850676
};

\end{axis}

\end{tikzpicture}
        \caption{Latency (AF and AO).}
        \label{fig:perc_lat_agg}
    \end{subfigure}
    \begin{subfigure}[b]{.4\linewidth}
        \centering
\begin{tikzpicture}

\begin{axis}[
width=\sfwidth,
height=\sfheight,
legend cell align={left},
legend style={at={(0.995,0.99)}, anchor=north east, fill opacity=0.8, legend columns=2, draw opacity=1, text opacity=1, font=\tiny, draw=white!80!black},
tick align=outside,
tick pos=left,
x grid style={white!69.0196078431373!black},
xlabel={$p_s$},
xmin=0.9, xmax=1,
xtick style={color=black},
y grid style={white!69.0196078431373!black},
ylabel={$\frac{B}{M}$ (MHz)},
ymin=10, ymax=35,
ytick style={color=black},
xmajorgrids,
ymajorgrids
]
\addplot[thick, black, dashed, forget plot]
table{
0.9 6
1 6
};

\addplot [thick, color2, mark=o, mark options={solid}]
table {%
0.9 14.1051088767621
0.905 14.1664783457547
0.91 14.2325448780922
0.915 14.3070023586975
0.92 14.3791757497502
0.925 14.4593281433173
0.93 14.5457498382962
0.935 14.6349823884224
0.94 14.7375098216803
0.945 14.8454829192762
0.95 14.9681690791311
0.955 15.0981175120222
0.96 15.2460482637842
0.965 15.4179171624717
0.97 15.6122196658076
0.975 15.8463501426533
0.98 16.1434813264916
0.985 16.5219011430087
0.99 17.0900169425588
0.995 18.183922421507
};

\addplot [thick, dashdotted, color2, mark=x, mark options={solid}]
table {%
0.9 14.775131634818
0.905 14.8624799675406
0.91 14.9561352348841
0.915 15.054087874685
0.92 15.1590131519467
0.925 15.2706496156585
0.93 15.3899314552404
0.935 15.5177277632867
0.94 15.6577194109906
0.945 15.8063296711503
0.95 15.973560444185
0.955 16.1610219338825
0.96 16.3678645691745
0.965 16.6036959159786
0.97 16.8770777860641
0.975 17.204327680823
0.98 17.6096121600148
0.985 18.1465039366806
0.99 18.9289150147099
0.995 20.4312385811133
};

\addplot [thick, color4, mark=o, mark options={solid}]
table {%
0.9 19.7504705032258
0.905 19.7971515760834
0.91 19.9019273943098
0.915 20.0240136962083
0.92 20.1009082816213
0.925 20.169062973354
0.93 20.3494953959904
0.935 20.4144224775559
0.94 20.6050834600759
0.945 20.6694931646137
0.95 20.9228371310181
0.955 21.068007936042
0.96 21.2292361318917
0.965 21.5084069467507
0.97 21.7802638346775
0.975 22.1075516686214
0.98 22.4358259547068
0.985 22.9848902423552
0.99 23.7886052970467
0.995 25.3660210656816
};

\addplot [thick, dashdotted, color4, mark=x, mark options={solid}]
table {%
0.9 21.140455788271
0.905 21.2624377179091
0.91 21.4024600081664
0.915 21.5574204357352
0.92 21.6924242186858
0.925 21.8725967691702
0.93 22.0696170806659
0.935 22.2319269981601
0.94 22.4010810584866
0.945 22.6379365638365
0.95 22.884067034119
0.955 23.1758946363667
0.96 23.4453675040108
0.965 23.8162259162759
0.97 24.1931731398537
0.975 24.6774741144437
0.98 25.28529535137
0.985 26.1029657505012
0.99 27.288437860862
0.995 29.6236792001275
};

\addplot [thick, color6, mark=o, mark options={solid}]
table {%
0.9 22.5728464358189
0.905 22.7885219636813
0.91 22.6909061779268
0.915 22.8636982575864
0.92 22.9269142221898
0.925 23.02375661174
0.93 23.2694310657861
0.935 23.2297445939395
0.94 23.142874948327
0.945 23.5549642536369
0.95 23.7020933320144
0.955 23.6050929487702
0.96 23.924935797366
0.965 24.2751372134983
0.97 24.405409988921
0.975 24.7014679158882
0.98 25.0512166862822
0.985 25.3883192166507
0.99 26.4657844907747
0.995 27.7176724457888
};

\addplot [thick, dashdotted, color6, mark=x, mark options={solid}]
table {%
0.9 24.3282473327281
0.905 24.3554878085025
0.91 24.4721346457791
0.915 24.7091061048078
0.92 24.7995266493474
0.925 25.0983227468878
0.93 25.0871882105453
0.935 25.3906084984181
0.94 25.4969581314748
0.945 25.7587174910972
0.95 26.1430236079887
0.955 26.3353348243823
0.96 26.6626107129973
0.965 27.2155061201159
0.97 27.5480793116527
0.975 28.0289547215271
0.98 28.8305304222325
0.985 29.49264290834
0.99 30.9059372935946
0.995 33.5352124136728
};

\end{axis}

\end{tikzpicture}
        \caption{Assigned bandwidth (IF and IO).}
        \label{fig:perc_sch_ind}
    \end{subfigure}
    \begin{subfigure}[b]{.4\linewidth}
        \centering
\begin{tikzpicture}

\begin{axis}[
width=\sfwidth,
height=\sfheight,
legend cell align={left},
legend style={at={(0.995,0.99)}, anchor=north east, fill opacity=0.8, legend columns=2, draw opacity=1, text opacity=1, font=\tiny, draw=white!80!black},
tick align=outside,
tick pos=left,
x grid style={white!69.0196078431373!black},
xlabel={$p_s$},
xmin=0.9, xmax=1,
xtick style={color=black},
y grid style={white!69.0196078431373!black},
ylabel={$\frac{B}{M}$ (MHz)},
ymin=10, ymax=35,
ytick style={color=black},
xmajorgrids,
ymajorgrids
]
\addplot[thick, black, dashed, forget plot]
table{
0.9 6
1 6
};

\addplot [thick, color2, mark=triangle, mark options={solid}]
table {%
0.9 12.9820328507458
0.905 13.0030727245937
0.91 13.0276696634261
0.915 13.0503073494248
0.92 13.0759880992014
0.925 13.1028879410104
0.93 13.1318626497853
0.935 13.1621301171201
0.94 13.1948227369744
0.945 13.2315767838548
0.95 13.2707782835075
0.955 13.3078348805846
0.96 13.3562217778458
0.965 13.4093524618822
0.97 13.4631613791774
0.975 13.54054624539
0.98 13.6230397794713
0.985 13.7379012851838
0.99 13.8892727005909
0.995 14.1890248177907
};

\addplot [thick, dashdotted, color2, mark=square, mark options={solid}]
table {%
0.9 13.2475548737121
0.905 13.2786579614138
0.91 13.3130077364417
0.915 13.3474246277495
0.92 13.3842358644192
0.925 13.4228432000811
0.93 13.4632324937478
0.935 13.5075502900404
0.94 13.5534255152916
0.945 13.6046420068009
0.95 13.6582161919647
0.955 13.7187356681771
0.96 13.7845535180595
0.965 13.8602480513649
0.97 13.9455380518029
0.975 14.0484998485219
0.98 14.169404169706
0.985 14.3282472371732
0.99 14.5554290365698
0.995 14.9815784870667
};

\addplot [thick, color4, mark=triangle, mark options={solid}]
table {%
0.9 19.874904700395
0.905 19.907175549227
0.91 19.9520057761903
0.915 19.9750341051514
0.92 20.0314285972253
0.925 20.0538887582158
0.93 20.0947067857526
0.935 20.1557174237759
0.94 20.2031847725953
0.945 20.264978213463
0.95 20.3091880460812
0.955 20.3639821394786
0.96 20.4490736592658
0.965 20.545845988262
0.97 20.6273726577019
0.975 20.7463479876166
0.98 20.8648040954708
0.985 21.0219025740447
0.99 21.291066375667
0.995 21.7626752438011
};

\addplot [thick, dashdotted, color4, mark=square, mark options={solid}]
table {%
0.9 20.3448570986024
0.905 20.3767337516832
0.91 20.4281885957729
0.915 20.49035196449
0.92 20.5582408068584
0.925 20.6071843637178
0.93 20.6654435690909
0.935 20.7338584465064
0.94 20.8010613047755
0.945 20.889748911631
0.95 20.9700631089721
0.955 21.0478479465522
0.96 21.1719219991435
0.965 21.3047295249689
0.97 21.449212892551
0.975 21.5851658656209
0.98 21.7992667721086
0.985 22.010592238363
0.99 22.3566097211459
0.995 23.0108223530774
};

\addplot [thick, color6, mark=triangle, mark options={solid}]
table {%
0.9 22.7443999393286
0.905 22.798467727251
0.91 22.8662715499928
0.915 22.861404664782
0.92 22.8790969997258
0.925 22.9255668380262
0.93 23.0009717027779
0.935 23.092608241063
0.94 23.1167633900975
0.945 23.1828420563385
0.95 23.2357920918416
0.955 23.3294027503686
0.96 23.3755815156597
0.965 23.4876599228492
0.97 23.582187871545
0.975 23.6775410246649
0.98 23.7966026946227
0.985 24.0802243021268
0.99 24.2685650009766
0.995 24.8404993447576
};

\addplot [thick, dashdotted, color6, mark=square, mark options={solid}]
table {%
0.9 23.3323628396177
0.905 23.3856385367434
0.91 23.4227179257729
0.915 23.4547805410565
0.92 23.5763340182909
0.925 23.6130461117999
0.93 23.7051753302545
0.935 23.7891460280281
0.94 23.8522803514425
0.945 23.9083766463751
0.95 24.0696162998359
0.955 24.1961371602794
0.96 24.2841629210064
0.965 24.4540148930361
0.97 24.5070373183305
0.975 24.7406419634265
0.98 24.9792906216519
0.985 25.2752124643757
0.99 25.7105756348158
0.995 26.4075515466479
};

\end{axis}

\end{tikzpicture}
        \caption{Assigned bandwidth (AF and AO).}
        \label{fig:perc_sch_agg}
    \end{subfigure}
    \caption{Average and worst-case percentiles of the latency and assigned bandwidth as a function of the quantile $p_s$.}
    \label{fig:multi_perc}
\end{figure*}

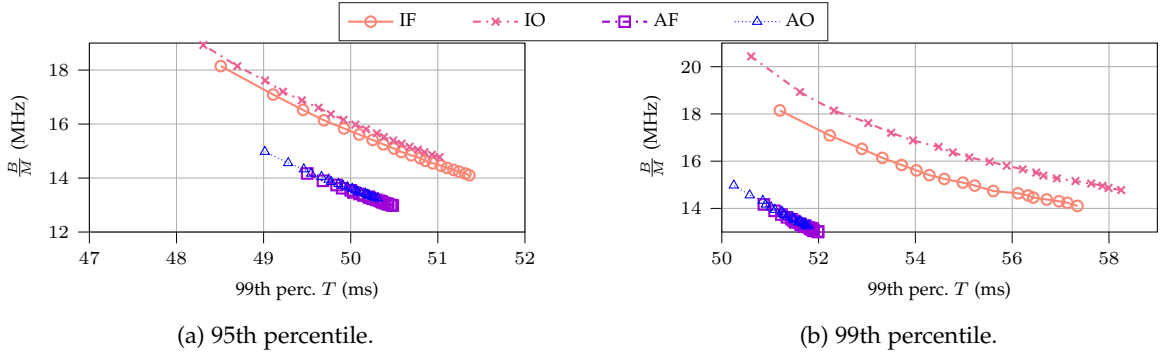
\begin{figure*}[t!]
    \centering
    \begin{subfigure}[b]{\linewidth}
        \centering
\begin{tikzpicture}

\begin{axis}[
    width=0,
    height=0,
    at={(0,0)},
    scale only axis,
    xmin=0,
    xmax=0,
    xtick={},
    ymin=0,
    ymax=0,
    ytick={},
    axis background/.style={fill=white},
    legend style={legend cell align=center, align=center, draw=white!15!black, font=\scriptsize, at={(0, 0)}, anchor=center, legend columns=6, /tikz/every even column/.append style={column sep=2em}}  
]
\addplot [thick, color2, mark=o, mark options={solid}]
table {%
0 1
};
\addlegendentry{\gls{if}}

\addplot [thick, dashdotted, color3, mark=x, mark options={solid}]
table {%
0 1
};
\addlegendentry{\gls{io}}

\addplot [thick, dashdotted, color5, mark=square, mark options={solid}]
table {%
0 1
};
\addlegendentry{\gls{af}}

\addplot [densely dotted, color6, mark=triangle, mark options={solid}]
table {%
0 1
};
\addlegendentry{\gls{ao}}

\end{axis}

\end{tikzpicture}
    \end{subfigure}
    \\
    \begin{subfigure}[b]{.45\linewidth}
        \centering
\begin{tikzpicture}

\begin{axis}[
width=\sfwidth,
height=\sfheight,
legend cell align={left},
tick align=outside,
tick pos=left,
x grid style={white!69.0196078431373!black},
xlabel={99th perc. $T$ (ms)},
xmin=47, xmax=52,
xtick style={color=black},
y grid style={white!69.0196078431373!black},
ylabel={$\frac{B}{M}$ (MHz)},
ymin=12, ymax=19,
ytick style={color=black},
xmajorgrids,
ymajorgrids
]
\addplot [thick, color2, mark=o, mark options={solid}]
table {%
51.3620486331896 14.1052311371433
51.3161839453036 14.1670263488114
51.2439471326641 14.2383271328271
51.1793129313552 14.3020487684368
51.1044301217189 14.3774372099466
51.0357048025092 14.4584597998358
50.9422481318062 14.5475594828726
50.8569093469492 14.6388497871294
50.7976666001117 14.7375774064444
50.693887576886 14.8452225551073
50.5806899201186 14.9659899796216
50.4973950563452 15.0923738205185
50.375596946191 15.2472297110134
50.2515620248649 15.4108895251749
50.0975604394016 15.6105989369109
49.9231485190025 15.8379956908828
49.6932382987074 16.1391144159407
49.4527560388415 16.521086251831
49.1097261004275 17.0900050920647
48.5083394359572 18.1475244672248
};
\addplot [thick, dashdotted, color3, mark=x, mark options={solid}]
table {%
51.0206676286993 14.7751260910105
50.9317259612854 14.8635907503953
50.8465400070246 14.9555663825669
50.7714826448298 15.053417136658
50.6764350215209 15.1578740100141
50.5853167320269 15.270438192577
50.4880073587736 15.3875584379298
50.3821647650442 15.518523379128
50.3007307603073 15.6566415714683
50.1758673458617 15.8068025989931
50.0531231284218 15.9764161426939
49.9187806703927 16.1603981449686
49.7722278057928 16.3674106526416
49.6307257252078 16.6041747140843
49.4390170468785 16.8768771777898
49.2218978631222 17.1993569059786
49.0164538744807 17.6135268039248
48.6978178607683 18.144064880031
48.3079575413066 18.9288943871561
47.653382354447 20.4351087807048
};
\addplot [thick, dashdotted, color5, mark=square, mark options={solid}]
table {%
50.4830795511061 12.9807099837023
50.4632835445736 13.0016005743504
50.4305281293464 13.0253574493991
50.4168117921669 13.0511633867428
50.3883186109781 13.0730561615835
50.3596660379222 13.1036315711636
50.3526685999443 13.1305659724709
50.3261757700866 13.1591921930647
50.2930442981279 13.1930728748796
50.2541369172956 13.2319570776296
50.2179327353133 13.2691443562924
50.2036864265206 13.307263714365
50.1469671657877 13.3524322937705
50.1014112296256 13.4083805511759
50.0306691948 13.4691299204728
49.9997244226535 13.5441092573678
49.9017076021453 13.6223017036291
49.8381161964863 13.7402930066579
49.682004906654 13.8983963030025
49.5008304400145 14.1690821091894
};
\addplot [densely dotted, color6, mark=triangle, mark options={solid}]
table {%
50.310774081281 13.2491051255639
50.2758211992148 13.2802269397099
50.2490537066062 13.312968101444
50.2136460591781 13.3476803737536
50.1891913148068 13.3852450847306
50.169385405553 13.4212250184653
50.1161286575038 13.464935196621
50.0767829000677 13.5057730142912
50.0522775167017 13.551203892903
50.0172444940452 13.6067590016229
49.9639716862194 13.6600486974365
49.9103827245143 13.7198769235474
49.859264647173 13.7849679743568
49.7737773427692 13.8620530525771
49.7469621770078 13.945255158612
49.6649854539111 14.046409518438
49.5460089142929 14.1671382617925
49.461371257344 14.3291421233145
49.2826004775638 14.5593394724833
49.0143654244395 14.9758826486193
};
\end{axis}

\end{tikzpicture}
        \caption{95th percentile.}
        \label{fig:pareto_95}
    \end{subfigure}
    \begin{subfigure}[b]{.45\linewidth}
        \centering
\begin{tikzpicture}

\begin{axis}[
width=\sfwidth,
height=\sfheight,
legend cell align={left},
tick align=outside,
tick pos=left,
x grid style={white!69.0196078431373!black},
xlabel={99th perc. $T$ (ms)},
xmin=50, xmax=59,
xtick style={color=black},
y grid style={white!69.0196078431373!black},
ylabel={$\frac{B}{M}$ (MHz)},
ymin=13, ymax=21,
ytick style={color=black},
xmajorgrids,
ymajorgrids
]
\addplot [thick, color2, mark=o, mark options={solid}]
table {%
57.35003457094 14.1052311371433
57.1404629596521 14.2383271328271
56.9693645424937 14.3020487684368
56.7136674837669 14.3774372099466
56.43583403921 14.4584597998358
56.3311400359943 14.5475594828726
56.1194190218212 14.6388497871294
55.6097289471514 14.7375774064444
55.2267331206301 14.9659899796216
54.9807821470537 15.0923738205185
54.5980793808485 15.2472297110134
54.2886393592008 15.4108895251749
54.021297720302 15.6105989369109
53.7163893920016 15.8379956908828
53.318584884566 16.1391144159407
52.8952027709983 16.5210862518311
52.2308807901911 17.0900050920647
51.2007992508688 18.1475244672248
};
\addplot [thick, dashdotted, color3, mark=x, mark options={solid}]
table {%
58.2515617456766 14.7751260910105
57.99188998225 14.8635907503953
57.8657368968592 14.9555663825669
57.6326946379865 15.053417136658
57.3066637650127 15.1578740100141
56.9239246522627 15.270438192577
56.6443452465857 15.3875584379298
56.4980110857789 15.518523379128
56.2184611789753 15.6566415714683
55.8867814743822 15.8068025989931
55.5331162580053 15.9764161426939
55.1133598411166 16.1603981449686
54.7718517494505 16.3674106526416
54.4812884543588 16.6041747140843
53.9518895443168 16.8768771777898
53.4990422107278 17.1993569059786
53.0271672790707 17.6135268039248
52.3171376518151 18.144064880031
51.6149888386018 18.9288943871561
50.6066173278435 20.4351087807048
};
\addplot [thick, dashdotted, color5, mark=square, mark options={solid}]
table {%
52.0444076028783 12.9807099837023
51.9846941088037 13.0016005743504
51.9907383947708 13.0253574493991
51.8946657101541 13.0511633867428
51.8875999623072 13.0730561615835
51.8701541532783 13.1036315711636
51.8278743860464 13.1305659724709
51.8691890646849 13.1591921930647
51.7981141473131 13.1930728748796
51.6945287574494 13.2319570776296
51.7272263324457 13.2691443562924
51.6352453026454 13.307263714365
51.62043423118 13.3524322937705
51.5181576283083 13.4083805511759
51.484575214819 13.4691299204728
51.4397558853167 13.5441092573678
51.3456103222482 13.622301703629
51.2313212730599 13.7402930066579
51.0904946755996 13.8983963030025
50.8700268666594 14.1690821091894
};
\addplot [densely dotted, color6, mark=triangle, mark options={solid}]
table {%
51.7454415256486 13.2491051255639
51.7980613034372 13.2802269397099
51.7381477735549 13.312968101444
51.7070488407749 13.3476803737536
51.6674890777105 13.3852450847306
51.5841861698721 13.4212250184653
51.6226499613519 13.464935196621
51.4823418094633 13.5057730142912
51.4936150469121 13.551203892903
51.4622904745843 13.6067590016229
51.3553180644997 13.6600486974365
51.2854037193513 13.7198769235474
51.2723432775462 13.7849679743568
51.2417372301475 13.8620530525771
51.097611179905 13.945255158612
51.0181085462598 14.046409518438
50.9117451028985 14.1671382617925
50.8471264853868 14.3291421233145
50.5757769051244 14.5593394724833
50.2487438029671 14.9758826486193
};
\end{axis}

\end{tikzpicture}
        \caption{99th percentile.}
        \label{fig:pareto_99}
    \end{subfigure}
    \caption{Pareto curves for the performance of the slicing schemes.}
    \label{fig:pareto}
\end{figure*}

We then analyze the per-user results more in depth by considering a scenario with $M=6$ users, whose detailed parameters are given in Table~\ref{tab:scenario}: each user has a different video bit rate $R_m$ and a different spectral efficiency, but they have a common frame rate $\varphi$ and are synchronized. As we did above, we consider a predictive slicing orchestrator using the general model obtained by performing the regression over all traces. The videos also have a random offsets, so even users with the same content have independent traces. Fig.~\ref{fig:per_user} shows the latency and assigned bandwidth for the 6 users, presenting some interesting patterns. As Fig.~\ref{fig:lat_peruser} shows, the latency for all users is the same when using the aggregated slicing schemes: this is an effect of using a need-based scheduler when allocating resources inside the slice. On the other hand, latency is different when using individual slicing, with users 5 and 6 having a lower latency violation probability. On the other hand, the bandwidth assigned to each user, shown in Fig.~\ref{fig:sch_peruser}, is similar for all schemes, with the aggregated ones having a slightly lower resource utilization. This is due to the fact that the average required bandwidth $\bar{B}_m=\frac{R_m}{\eta_m}$, whose value for each user is given in Table~\ref{tab:scenario}, is the main factor in assigning bandwidth resources, both with individual and aggregated schemes: we can easily see that users 1, 2, and 5, who have the lowest values of $\bar{B}_m$, are also assigned the least bandwidth by the slicing schemes. Interestingly, user 5 is actually the one with the lowest bandwidth, although users 1 and 2 have lower values of $\bar{B}_m$: this is due to the higher relative variations in the video traces at lower bit rates, as can be easily seen from Fig.~\ref{fig:box_gen}, so that in order to meet the $p_s=0.95$ requirement, the slicing schemes have to overprovision more for those users.

We can also look at the performance as a function of $p_s$ in the scenario given in Table~\ref{tab:scenario}, as we did for the single-user case. The results are shown in Fig.~\ref{fig:multi_perc}: it is easy to see that aggregated schemes manage to maintain a much lower latency in the worst case with a lower total bandwidth, as the individual schemes have either a latency higher than the 50~ms threshold (for lower values of $p_s$) or a far higher resource utilization (for higher values of $p_s$). Interestingly, the performance difference between \gls{fdma} and \gls{ofdma} is negligible for the aggregated schemes: the prediction errors of a user are effectively compensated by the other users, enhancing the overall performance.

We can further analyze the performance difference between the schemes by plotting their Pareto curves. Pareto curves are useful to show two-dimensional performance metrics which need to be traded against one another: the curve includes all points at the edge of the achievable performance region, i.e., points for which it is impossible to improve one metric without making the other worse. In our case, the two metrics are the \gls{mtp} latency and the bandwidth: ideally, we would like both to be as low as possible, and the Pareto curve is another way of showing the trade-off we discussed above.

If we define the performance of a slicing scheme $g$ in terms of latency and bandwidth as $q_g(p_s)=(T,B)$, we can say that $p_s$ \emph{dominates} $p_s'$, and we write $p_s \succ p_s'$ where $p_s$ has a better performance for both metrics, i.e.,
\begin{equation}
    T(g,p_s)<T(g,p_s') \wedge B(g,p_s)<B(g,p_s').
\end{equation}
We can then define the Pareto curve $\mathcal{P}_g$ as the set of points that are not dominated by any other point:
\begin{equation}
    \mathcal{P}_g \triangleq \left\{q_g(p_s), \forall p_s \mid \nexists p_s': p_s' \succ p_s\right\}.
\end{equation}
Fig.~\ref{fig:pareto} shows the Pareto curves for the 4 schemes, considering $p_s\in[0.9,0.995]$. The two plots show the performance in terms of the average assigned bandwidth per user and the 95th and 99th percentiles of latency, and confirm our analysis: the aggregated schemes can significantly outperform the individual ones, with a bandwidth reduction of more than 10\% to obtain the same latency performance at the 95th percentile and more than 20\% at the 99th percentile. We can also note that, while the difference between \gls{if} and \gls{io} is relatively small for the 95th percentile, it grows for the 99th percentile, as taking the worst case highlights the limits of the \gls{fdma} approach. On the other hand, the difference between \gls{af} and \gls{ao} is negligible. In addition to significantly improving the performance, choosing an aggregated scheme is also computationally simpler, as the slicing algorithm will only need to allocate resources to a single \gls{vr} slice and not to each individual user.

\section{Conclusions}
\label{sec:conclusions}

This work aims at closing a gap in the literature on traffic source modeling: there are several analyses for passive streaming, both 2D and in immersive setups with \glsfirstplural{hmd}, and some for live gaming traffic in 2D, but none for interactive \gls{vr} with strict latency requirements and \textit{quasi}-\gls{cbr} encoding. We analyzed live captures from a setup we devised, publishing both the dataset and the code for the analysis, and presented the performance of two regression models. The first part of our discussion analyzes the prediction models, determining the necessary memory in the linear regression, the residual distribution, and the correctness of the linear model. The prediction models are simple and flexible, as they generalize extremely well across different traces and bit rate settings: this means that a shared pre-trained model can be used with good performance across different video content types and bit rate levels. 

We then showed a simple \glsfirst{ns} scenario, which highlights the importance of the trade-off between resource efficiency and \gls{qoe}. This is a first step towards fully designing an \gls{ns} system able to satisfy the stringent \gls{qos} requirements of \gls{xr} applications also in critical scenarios, e.g., in industrial settings, in which the consequences of network failures are not only discomfort and nausea for the user, but also significant delays in production and even safety hazards. The results we obtained show a significant trade-off between resource efficiency and \gls{mtp} latency guarantees, which can be improved significantly if multiple \gls{vr} users are put together in the same slice, sharing \glsentryfull{ran} resources using a fair scheduler.

There are several additional analyses and opportunities for future work, that can be divided in two main directions.
The first potential avenue of research is a wider characterization, with different encoding parameters and even different encoders, and considering different applications, going beyond simple \gls{vr} games to include the industrial and commercial use cases we mentioned above, and a wider set of subjects. The traces should also integrate a record of the head movements of the users, as they correspond to shifts in the point of view of the \gls{vr} headset and are expected to be strongly correlated with frame size changes.

The other challenge is to actually design slicing schemes and scheduling algorithms able to take into account the nature of the traffic and accommodate it, efficiently exploiting the prediction and adapting to the peculiarities of different communication technologies or even multiple independent links. The use of packet-level coding to protect the stream from link failures and deep fading events, can be promising avenues to design a solid framework to support \gls{xr} in mission-critical scenarios. The study of these techniques at all levels of the communication stack, simulating connection impairments in repeatable conditions through a full-stack network simulator, is our first priority in the ongoing work on this subject.

\appendix
\subsection*{Aggregated traffic distribution}

Given that the frame size distribution for a single user is known and well defined, we can derive the distribution of the total required bandwidth $B^*_f(k)$ for the $M$ users in closed form. In the following, we will consider the case in which $S=1$, but the calculation can be trivially extended to the case with $S>1$.
Specifically, the bandwidth required to deliver the next frame for all the $M$ users is defined in~\cref{eq:B_req_AF} as a weighted sum of Laplace random variable:
\begin{equation}
    B_{f}^*(k) = \sum_{m=1}^M\eta_m^{-1} F^{(m)}(k).
\end{equation}
In the following, we omit the index $k$ for the sake of readability.
The user spectral efficiencies $\eta_m^{-1}$ are constant scalars which result in a scaling of the frame size distribution, i.e.,
\begin{equation}
    B_m = \eta_m^{-1} F^{(m)} \sim\mathrm{Laplace}\left(\alpha_m,\beta_m\right),
\end{equation}
where $\alpha_m = \frac{\mu_m}{\eta_m}$ and $\beta_m = \frac{b_m}{\eta_m}$.
The distribution of the sum of independent random variables is obtained as the convolution of the corresponding \glspl{pdf}, denoted by $p_m(x)$.
In the Laplace transform domain, this translates to a product of the transforms of the densities.
Given that the Laplace transform $\mathcal{L}[\cdot]$ of the Laplace distribution is
\begin{equation}
\begin{aligned}
    \mathcal{L}\left[p_m(x)\right](s) = &\mathcal{L}\left(\frac{1}{2\beta_m}e^{-\frac{|x-\alpha_m|}{\beta_m}},s\right)\\
    =& e^{-\alpha_m s} \frac{1}{1-\beta_m^2s^2},
\end{aligned}
\end{equation}
the transform of the \gls{pdf} $p(x)$ of the sum $B_f^*(k)$ is
\begin{equation}
    \mathcal{L}\left[p(x)\right](s)=e^{-s\sum\limits_{m=1}^M\alpha_m}\prod_{m=1}^M\frac{1}{1-\beta_m^2s^2}.
    \label{eq:sum_tf}
\end{equation}
In the case when the poles have multiplicity equal to $1$, the product can be represented with its partial function decomposition as
\begin{equation}
    \begin{aligned}
        \mathcal{L}\left[p(x)\right](s)= & e^{-s\sum\limits_{m=1}^M\alpha_m}\prod_{m=1}^M\frac{1}{1-\beta_m^2s^2} \\ 
        =& e^{-s\sum\limits_{m=1}^M\alpha_m}\sum_{m=1}^M\frac{\gamma_m}{1-\beta_m^2s^2},
    \end{aligned}
    \label{eq:prod}
\end{equation}
where the values of $\gamma_m$ can be obtained through the residue (or Heaviside cover-up) method, and are equal to
\begin{equation}
    \gamma_m =\frac{\beta_m^{2(M-1)}}{\prod_{n\ne m} \left( \beta_m^2-\beta_n^2\right)}.
\end{equation}
Thanks to the additive and the shift property of the Laplace transform, we get the \gls{pdf} of
\begin{equation}
    \begin{aligned}
        p\left(x-\sum_{m=1}^M\alpha_m\right)= & \mathcal{L}^{-1}\left[\mathcal{L}\left[p(x)\right](s)\right]\left(x-\sum_{m=1}^M\alpha_m\right) 
        \\ =&\sum_{m=1}^M \frac{\beta_m^{2(M-1)-1}e^{-\frac{|x|}{\beta_m}}}{2\prod_{n\ne m}\left( \beta_m^2-\beta_n^2\right)}.
    \end{aligned}
\end{equation}
We can then use this result to directly compute the quantile function $P_{1,\ldots,M}^{-1}(p_s|1,1,k)$. Extending the result to larger values of $S$ and multiple frames, we can simply plug in the function to obtain the \gls{af} and \gls{ao} slicing schemes.

The above calculation is correct in the case of simple poles. We can then briefly discuss the more general case of repeated poles.
Repeated poles can be present in \cref{eq:sum_tf} if the distributions of the bandwidth required by different users have the same shape parameter $\beta$ (i.e., the ratio between the shape parameter $b_m$ and the user spectral efficiency $\eta_m$), regardless of the mean $\alpha$.
This may be the case when the same or similar applications are streamed to multiple, possibly static users in the same room, e.g., in the case of a \gls{vr} arena.
Thus, the distribution of the frame sizes, determined by the application, and the spectral efficiencies, determined by the propagation environment, may be the same for different users.
Namely, we can consider the most general case when there are $P\le M$ unique poles $\left\{\beta_1, \beta_2, \ldots, \beta_P\right\}$ of order $\left\{p_1, p_2, \ldots, p_P\right\}$, $\sum_{i=1}^P{p_i}=M$, i.e., the users are grouped in $P$ clusters with similar $\beta$ values.

In this case, the partial fraction representation is more involved than that in \cref{eq:prod}.
Taking into account the repeated poles, the product in \cref{eq:prod} can be rewritten as
\begin{equation}
    \begin{aligned}
        \prod_{m=1}^M\frac{1}{1-\beta_m^2s^2} = \prod_{i=1}^P\frac{1}{\left(1-\beta_i^2s^2\right)^{p_i}},
    \end{aligned}
    \label{eq:f_s}
\end{equation}
and can thus be decomposed as
\begin{equation}
    \prod_{i=1}^P\frac{1}{\left(1-\beta_i^2s^2\right)^{p_i}} = \sum_{h=1}^P \sum_{k=1}^{p_h}\frac{\gamma_{h,k}}{\left(1-\beta_h^2s^2\right)^{k}} = f(s).
\end{equation}
Several methods exist for determining the $\gamma$ vector.
The residue method offers a general solution:
\begin{equation}
    \gamma_{h,k} = \text{Res}\left(g_{h,k},\beta^2_h\right),
\end{equation}
where $\text{Res}(g_{h,k},\cdot)$ is the residue of the function
\begin{equation}
    g_{h,k}(s) = \left(1-\beta_h^2s^2\right)^{k-1} f(s),
\end{equation}
and can be computed as
\begin{equation}
    \begin{aligned}
        &\text{Res}\left(g_{h,k},\beta^2_h\right)=
        \\ 
        &=\frac{1}{(p_h-k)!}\left[ \frac{\dd^{p_h-k}}{\dd (s^2)^{p_h-k}}\left(1-\beta_h^2s^2\right)^{p_h}f(s)\right]_{s^2=\beta^2_h}.
    \end{aligned}
\end{equation}

\bibliographystyle{IEEEtran}
\bibliography{./bibl}

\end{document}